\documentclass[prd,twocolumn,groupedaddress,showpacs,nofootinbib]{revtex4}
\usepackage{graphicx}
\usepackage{dcolumn}
\usepackage{amssymb}
\usepackage{mathrsfs}
\usepackage{amsmath}
\usepackage{epsfig}
\usepackage[dvips]{color}
\usepackage{hhline}

\begin{document}

\title{From Dirac neutrino masses to baryonic and dark matter asymmetries}

\author{Pei-Hong Gu}
\email{peihong.gu@mpi-hd.mpg.de}

\affiliation{Max-Planck-Institut f\"{u}r Kernphysik, Saupfercheckweg
1, 69117 Heidelberg, Germany}

\begin{abstract}

We consider an $SU(3)'^{}_c\times SU(2)'^{}_L\times U(1)'^{}_Y$ dark
sector, parallel to the $SU(3)^{}_c\times SU(2)^{}_L\times
U(1)^{}_Y$ ordinary sector. The hypercharges, baryon numbers and
lepton numbers in the dark sector are opposite to those in the
ordinary sector. We further introduce three types of messenger
sectors: (i) two or more gauge-singlet Dirac fermions, (ii) two or
more $[SU(2)_L^{}\times SU(2)'^{}_L]$-bidoublet Higgs scalars, (iii)
at least one gauge-singlet Dirac fermion and at least one
$[SU(2)_L^{}\times SU(2)'^{}_L]$-bidoublet Higgs scalar. The lepton
number conserving decays of the heavy fermion singlet(s) and/or
Higgs bidoublet(s) can simultaneously generate a lepton asymmetry in
the $[SU(2)_L^{}]$-doublet leptons and an opposite lepton asymmetry
in the $[SU(2)'^{}_L]$-doublet leptons to account for the
cosmological baryon asymmetry and dark matter relic density,
respectively. The lightest dark nucleon as the dark matter particle
should have a mass about $5\,\textrm{GeV}$. By integrating out the
heavy fermion singlet(s) and/or Higgs bidoublet(s), we can obtain
three light Dirac neutrinos composed of the ordinary and dark
neutrinos. If a mirror discrete symmetry is further imposed, our
models will not require more unknown parameters than the traditional
type-I, type-II or type-I+II seesaw models.

\end{abstract}


\maketitle

\section{Introduction}

Various neutrino oscillation experiments have established the
phenomenon of massive and mixing neutrinos. To naturally understand
the smallness of neutrino masses, we can consider the seesaw
\cite{minkowski1977} extension of the $SU(3)^{}_c\times SU(2)^{}_L
\times U(1)^{}_Y$ standard model (SM). In the conventional seesaw
scenario \cite{minkowski1977,mw1980,flhj1989}, the neutrinos have a
Majorana nature which, however, has not been experimentally
verified. Alternatively, we can build some Dirac seesaw models
\cite{rw1983,mp2002,gh2006} to give the light Dirac neutrinos.
Usually, the Dirac seesaw needs more parameters than the Majorana
seesaw. In the Majorana or Dirac seesaw context, we can obtain the
cosmological baryon asymmetry through the leptogenesis
\cite{fy1986,mz1992,fps1995,hms2000,di2002,hs2004,hlnps2004,hrs2005,dnn2008,bd2009,gu2012}
or neutrinogenesis \cite{dlrw1999,tt2006,mp2002,gh2006} mechanism.
On the other hand, the dark matter relic density also indicates the
necessity of supplementing the SM. It is intriguing that the
baryonic and dark matter contribute comparable energy densities to
the present universe \cite{komatsu2010} although they have very
different properties. This coincidence can be elegantly explained if
the dark matter relic density is an asymmetry between the dark
matter and antimatter and its origin is related to the baryon
asymmetry. Such asymmetric dark matter can naturally appear in the
mirror universe models
\cite{ly1956,kop1966,pavsic1974,bk1982,glashow1986,flv1991,flv1992,abs1992,hodges1993,bdm1996,silagadze1997,cf1998,
mt1999,bcv2001,bb2001,bgg2001,iv2003,fv2003,berezhiani2004,foot2004,berezhiani2005,berezhiani2006,
bb2006-1,bb2006,acmy2009,dlnt2011,chly2012}.
There are also other ideas connecting the dark matter asymmetry to
the baryon asymmetry
\cite{nussinov1985,bcf1990,kaplan1992,dgw1992,kuzmin1998,kl2005,as2005,
clt2005,gsz2009,dmst2010,bdfr2011,mcdonald2011,hmw2010,dk2011,
frv2011,hms2011,kllly2011,gsv2011,fss2011,myz2012,idc2011,bpsv2011,cr2011,
as2012,ms2011,dm2012,fnp2012}.

In this paper we shall propose a common genesis of the Dirac
neutrino masses, the baryon asymmetry and the dark matter relic
density. Specifically, we consider an $SU(3)'^{}_c\times
SU(2)'^{}_L\times U(1)'^{}_Y$ dark sector, parallel to the
$SU(3)^{}_c\times SU(2)^{}_L \times U(1)^{}_Y$ ordinary sector. The
hypercharges, baryon numbers and lepton numbers in the dark sector
are opposite to those in the ordinary sector. Besides the ordinary
and dark sectors, there are three types of messenger sectors: (i)
two or more gauge-singlet Dirac fermions, (ii) two or more
$[SU(2)_L^{}\times SU(2)'^{}_L]$-bidoublet Higgs scalars, (iii) at
least one gauge-singlet Dirac fermion and at least one
$[SU(2)_L^{}\times SU(2)'^{}_L]$-bidoublet Higgs scalar. Through the
lepton number conserving decays of the heavy fermion singlet(s)
and/or Higgs bidoublet(s), we can simultaneously realize a lepton
asymmetry in the $[SU(2)_L^{}]$-doublet leptons and an opposite
lepton asymmetry in the $[SU(2)'^{}_L]$-doublet leptons to account
for the baryon asymmetry and the dark matter relic density,
respectively. The lightest dark nucleon with a determined mass about
$5\,\textrm{GeV}$ can serve as the dark matter particle. The dark
photon will become massive although the ordinary photon keeps
massless. The kinetic mixing between the $U(1)^{}_Y$ and
$U(1)'^{}_Y$ gauge fields can result in a testable scattering of the
dark nucleons off the ordinary nucleons. Furthermore, we can get a
tiny mass term between the ordinary and dark left-handed neutrinos
by integrating out the heavy fermion singlet(s) and/or Higgs
bidoublet(s). So, the ordinary and dark neutrinos can naturally form
three light Dirac neutrinos
\cite{flv1992,abs1992,berezhiani2005,bb2006-1}. Finally, we can
impose a mirror discrete symmetry to reduce the parameters. In this
case, our models will not contain additional parameters compared
with the traditional Majorana seesaw models.

\section{The ordinary and dark sectors}

We denote the ordinary quarks, leptons and scalars by
\begin{eqnarray}
&&\begin{array}{r}q_L^{}(\textbf{3},\textbf{2},+\frac{1}{3})
=\left[\begin{array}{r}u^{}_{L}\\
[1mm]
d^{}_{L}\end{array}\right],~d^{}_R(\textbf{3},\textbf{1},-\frac{2}{3})\,,
~u^{}_R(\textbf{3},\textbf{1},+\frac{4}{3})\,,\end{array}\nonumber\\
&&\begin{array}{r}l_L^{}(\textbf{1},\textbf{2},-1)
=\left[\begin{array}{r}\nu^{}_{L}\\
[1mm] e^{}_{L}\end{array}\right],~e^{}_R(\textbf{1},\textbf{1},-2)\,,\end{array}\nonumber\\
&&\begin{array}{r}\phi(\textbf{1},\textbf{2},-1)=\left[\begin{array}{l}\phi^{0}_{}\\
[1mm]
\phi^{-}_{}\end{array}\right],~\delta(\textbf{1},\textbf{1},-4)\,,\end{array}
\end{eqnarray}
where the first and second numbers in parentheses are the dimensions
of the $SU(3)^{}_c$ and $SU(2)^{}_L$ representations, while the
third ones are the $U(1)^{}_Y$ hypercharges $Y$. Accordingly, we
define the dark quarks, leptons and scalars as below,
\begin{eqnarray}
&&\begin{array}{r}q'^{}_L(\textbf{3},\textbf{2},-\frac{1}{3})
=\left[\begin{array}{r}d'^{}_{L}\\
[1mm]
u'^{}_{L}\end{array}\right],~d'^{}_R(\textbf{3},\textbf{1},+\frac{2}{3})\,,
~u'^{}_R(\textbf{3},\textbf{1},-\frac{4}{3})\,,\end{array}\nonumber\\
&&\begin{array}{r}l'^{}_L(\textbf{1},\textbf{2},+1)
=\left[\begin{array}{r}e'^{}_{L}\\
[1mm] \nu'^{}_{L}\end{array}\right],~e'^{}_R(\textbf{1},\textbf{1},+2)\,,\end{array}\nonumber\\
&&\begin{array}{r}\phi'(\textbf{1},\textbf{2},+1)=\left[\begin{array}{l}\phi'^{+}_{}\\
[1mm]
\phi'^{0}_{}\end{array}\right],~\delta'(\textbf{1},\textbf{1},+4)\,,\end{array}
\end{eqnarray}
where the first and second numbers in parentheses are the dimensions
of the $SU(3)'^{}_c$ and $SU(2)'^{}_L$ representations, while the
third ones are the $U(1)'^{}_Y$ hypercharges $Y'$. Like the
hypercharges, the baryon or lepton numbers of the dark fermions are
assumed opposite to those of the ordinary fermions.

We then write down the Lagrangian of the ordinary and dark sectors,
\begin{eqnarray}
\mathcal{L}_{}^{\textrm{OD}}=\mathcal{L}_{K}^{\textrm{OD}}+\mathcal{L}_{Y}^{\textrm{OD}}-V^{\textrm{OD}}_{}\,.
\end{eqnarray}
Here the index "OD" is the abbreviation of "ordinary-dark". The
kinetic terms include
\begin{eqnarray}
\label{kinetic} \mathcal{L}_K^{\textrm{OD}}&=&
(D_\mu^{}\phi)^\dagger_{}(D^\mu_{}\phi)+(D_\mu^{}\delta)^\dagger_{}(D^\mu_{}\delta)
+i\bar{q}_{L_i^{}}^{}\!\not\!\!D
q_{L_i^{}}^{}\nonumber\\
&&+i\bar{d}_{R_i^{}}^{} \!\not\!\!D
d_{R_i^{}}^{}+i\bar{u}_{R_i^{}}^{} \!\not\!\!D u_{R_i^{}}^{}
+i\bar{l}_{L_i^{}}^{} \!\not\!\!D l_{L_i^{}}^{}\nonumber\\
&&+i\bar{e}_{R_i^{}}^{} \!\not\!\!D e_{R_i^{}}^{}
+(D_\mu^{}\phi')^\dagger_{}(D^\mu_{}\phi')+(D_\mu^{}\delta')^\dagger_{}(D^\mu_{}\delta')
\nonumber\\
&&+i\bar{q}'^{}_{L_i^{}}\!\not\!\!D
q'^{}_{L_i^{}}+i\bar{d}'^{}_{R_i^{}} \!\not\!\!D
d'^{}_{R_i^{}}+i\bar{u}'^{}_{R_i^{}}\! \not\!\!D u'^{}_{R_i^{}}+ i
\bar{l}'^{}_{L_i^{}} \!\not\!\!D
l'^{}_{L_i^{}}\nonumber\\
&&+i\bar{e}'^{}_{R_i^{}} \!\not\!\!D
e'^{}_{R_i^{}}-\frac{1}{4}G_{\mu\nu}^{a}G^{a\mu\nu}_{}-\frac{1}{4}W_{\mu\nu}^{a}W^{a\mu\nu}_{}
\nonumber\\
&&-\frac{1}{4}B_{\mu\nu}^{}B^{\mu\nu}_{}-\frac{1}{4}G_{\mu\nu}'^{a}G'^{a\mu\nu}_{}
-\frac{1}{4}W'^{a}_{\mu\nu}W'^{a\mu\nu}_{}
\nonumber\\
&&-\frac{1}{4}B'^{}_{\mu\nu}B'^{\mu\nu}_{}-\frac{\epsilon}{2}B_{\mu\nu}^{}B'^{\mu\nu}_{}\,,
\end{eqnarray}
where the covariant derivatives are given by
\begin{eqnarray}
D_\mu^{}&=&\partial_\mu^{}-i g_1^{}\frac{Y}{2}
B_\mu^{}-ig_2^{}\frac{\tau_a^{}}{2}W_{\mu}^a\delta^{}_2-i
g_3^{}\frac{\lambda_a^{}}{2}G_\mu^a\delta^{}_3\nonumber\\
&&-i g'^{}_1\frac{Y'}{2}
B'^{}_\mu-ig'^{}_2\frac{\tau_a^{}}{2}W'^a_{\mu}\delta'^{}_{2}-i
g'^{}_3\frac{\lambda_a^{}}{2}G'^a_\mu\delta'^{}_{3}~~\textrm{with}\nonumber\\
&&
\begin{array}{l}\delta_{2}^{}=\left\{\begin{array}{ll}1&\textrm{for}~SU(2)_{L}^{}~\textrm{doublets}\,,\\
0&\textrm{for}~SU(2)_{L}^{}~\textrm{singlets}\,;\end{array}\right.\\
[5mm]
\delta_{3}^{}=\left\{\begin{array}{ll}1&\textrm{for}~SU(3)_{c}^{}~\textrm{triplets}\,,\\
0&\textrm{for}~SU(3)_{c}^{}~\textrm{singlets}\,;\end{array}\right.\\
[5mm]
\delta'^{}_{2}=\left\{\begin{array}{ll}1&\textrm{for}~SU(2)'^{}_{L}~\textrm{doublets}\,,\\
0&\textrm{for}~SU(2)'^{}_{L}~\textrm{singlets}\,;\end{array}\right.\\
[5mm]
\delta'^{}_{3}=\left\{\begin{array}{ll}1&\textrm{for}~SU(3)'^{}_{c}~\textrm{triplets}\,,\\
0&\textrm{for}~SU(3)'^{}_{c}~\textrm{singlets}\,.\end{array}\right.\end{array}
\end{eqnarray}
We also show the Yukawa interactions:
\begin{eqnarray}
\mathcal{L}_{Y}^{\textrm{OD}}&=&-(y_d^{})_{ij}^{}\bar{q}^{}_{L_i^{}}
\tilde{\phi} d^{}_{R_j^{}} -(y_u^{})_{ij}^{}\bar{q}^{}_{L_i^{}} \phi
u^{}_{R_j^{}}\nonumber\\
&& -(y_e^{})_{ij}^{}\bar{l}^{}_{L_i^{}} \tilde{\phi} e^{}_{R_j^{}}
-(y_{\delta}^{})_{ij}^{}\delta\bar{e}^{}_{R_i^{}} e^c_{R_j^{}}\nonumber\\
&&- (y^{}_{d'})_{ij}^{}\bar{q}'^{}_{L_i^{}} \tilde{\phi}'
d'^{}_{R_j^{}} -(y^{}_{u'})_{ij}^{}\bar{q}'^{}_{L_i^{}} \phi'
u'^{}_{R_j^{}}\nonumber\\
&&-(y^{}_{e'})_{ij}^{} \bar{l}'^{}_{L_i^{}} \tilde{\phi}'
e'^{}_{R_j^{}} -(y_{\delta'}^{})_{ij}^{}\delta'\bar{e}'^{}_{R_i^{}}
e'^c_{R_j^{}}+\textrm{H.c.} \,,
\end{eqnarray}
and the scalar potential:
\begin{eqnarray}
V_{}^{\textrm{OD}}&=&\mu_\phi^2 \phi^\dagger_{}\phi
+\lambda_\phi^{}(\phi^\dagger_{}\phi)^2_{}+\mu_\delta^2
\delta^\dagger_{}\delta
+\lambda_\delta^{}(\delta^\dagger_{}\delta)^2_{}\nonumber\\
&&+2\lambda_{\phi\delta}^{}\phi^\dagger_{}\phi\delta^\dagger_{}\delta+\mu_{\phi'}^2
\phi'^\dagger_{}\phi'
+\lambda_{\phi'}^{}(\phi'^\dagger_{}\phi')^2_{}\nonumber\\
&&+\mu_{\delta'}^2 \delta'^\dagger_{}\delta'
+\lambda_{\delta'}^{}(\delta'^\dagger_{}\delta')^2_{}
+2\lambda_{\phi'\delta'}^{}\phi'^\dagger_{}\phi'\delta'^\dagger_{}\delta'\nonumber\\
&& +2\lambda_{\phi\phi'}^{}\phi^\dagger_{}\phi\phi'^\dagger_{}\phi'
+2\lambda_{\phi\delta'}^{}\phi^\dagger_{}\phi\delta'^\dagger_{}\delta'\nonumber\\
&&+2\lambda_{\delta\phi'}^{}\delta^\dagger_{}\delta\phi'^\dagger_{}\phi'
+2\lambda_{\delta\delta'}^{}\delta^\dagger_{}\delta\delta'^\dagger_{}\delta'\,.
\end{eqnarray}

The symmetry breaking pattern is expected to be
\begin{eqnarray}
SU(3)^{}_c\times SU(2)^{}_L\times U(1)^{}_Y&\longrightarrow&
SU(3)^{}_c\times U(1)^{}_{em}\,,\nonumber\\
SU(3)'^{}_c\times SU(2)'^{}_L\times U(1)'^{}_Y&\longrightarrow&
SU(3)'^{}_c\times U(1)'^{}_{em}\nonumber\\
&\longrightarrow& SU(3)'^{}_c\,.
\end{eqnarray}
For this purpose, the ordinary and dark scalars should develop their
vacuum expectation values (VEVs) as below,
\begin{eqnarray}
\langle\phi\rangle&=&\left[\begin{array}{c}\langle\phi^0_{}\rangle\\
[2mm] 0\end{array}\right]\simeq 174\,\textrm{GeV}\,,~~\langle\delta\rangle=0\,,\nonumber\\
\langle\phi'\rangle&=&\left[\begin{array}{c}0\\
[2mm]
\langle\phi'^{0}_{}\rangle\end{array}\right]\,,~~\langle\delta'\rangle\neq
0\,.
\end{eqnarray}
This means the ordinary photon will keep massless while the dark
photon will become massive.

It is straightforward to read the fermion masses in the ordinary
sector,
\begin{eqnarray}
\mathcal{L}&\supset&-m^{}_{d}\bar{d}^{}_L d^{}_R
-m^{}_{u}\bar{u}^{}_L u^{}_R -m^{}_{e}\bar{e}^{}_L e^{}_R +\textrm{H.c.} ~~\textrm{with}\nonumber\\
&&m^{}_d =y^{}_d\langle\phi\rangle\,,~m^{}_u
=y^{}_u\langle\phi\rangle\,,~m^{}_e=y^{}_e\langle\phi\rangle\,,
\end{eqnarray}
and the fermion masses in the dark sector,
\begin{eqnarray}
\label{dfmass} \mathcal{L}&\supset&-m^{}_{d'}\bar{d}'^{}_R d'^{}_L
-m^{}_{u'}\bar{u}'^{}_R u'^{}_L -m^{}_{e'}\bar{e}'^{}_R
e'^{}_L-\delta m^{}_{e'}\bar{e}'^{}_R e'^{c}_R\nonumber\\
&& +\textrm{H.c.} ~~\textrm{with}~~m^{}_{d'}
=y^{}_{d'}\langle\phi'\rangle\,, ~m^{}_{u'} =
y^{}_{u'}\langle\phi'\rangle\,,\nonumber\\
&&m^{}_{e'} = y^{}_{e'}\langle\phi'\rangle\,,~\delta
m^{}_{e'}=y^{}_{\delta'}\langle\delta'\rangle\,.
\end{eqnarray}
The dark charged leptons should be the so-called quasi-Dirac
fermions for $\delta m_{e'}^{}\ll m_{e'}^{}$. In the ordinary
sector, the quark masses $m_u^{}$ and $m_d^{}$ are much smaller than
the hadronic scale $\Lambda_{\textrm{QCD}}^{}$ so that they can only
have a negligible contribution to the nucleon masses,
\begin{eqnarray}
m_p^{}\simeq m_n^{}\simeq 1\,\textrm{GeV}=m_{N}^{}\,.
\end{eqnarray}
In the dark sector, the quark masses $m_{u'}^{}$ and $m_{d'}^{}$ may
be sufficiently larger than the hadronic scale
$\Lambda^{}_{\textrm{QCD}'}$. In this case, the dark nucleon masses
can approximately equal the sum of the dark quark masses,
\begin{eqnarray}
m_{p'}^{}=2m_{u'}^{}+m_{d'}^{}\,,~~ m_{n'}^{}=2m_{d'}^{}+
m_{u'}^{}\,.
\end{eqnarray}

As we will demonstrate in the following, our completed models also
contain a messenger sector. We will refer to
$\mathcal{L}_{}^{\textrm{M}}$ (with the index "M" being the
abbreviation of "messenger") as the Lagrangian involving the
messenger fields. Note that our models will not have any baryon or
lepton number violating interactions except the $SU(2)^{}_L$ and
$SU(2)'^{}_L$ sphaleron processes \cite{thooft1976,krs1985}.

\section{The model with gauge-singlet Dirac fermions}

\begin{figure}
\vspace{5.0cm} \epsfig{file=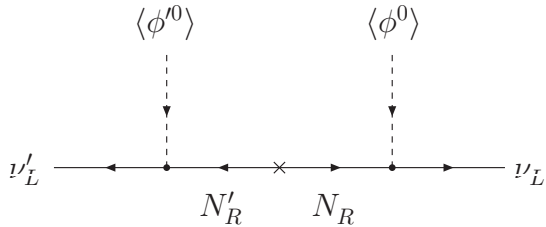, bbllx=5cm, bblly=6.0cm,
bburx=15cm, bbury=16cm, width=8.5cm, height=8.5cm, angle=0, clip=0}
\vspace{-10cm} \caption{\label{DiracI} The type-I Dirac seesaw for
generating the masses between the ordinary left-handed neutrinos
$\nu_L^{}$ and the dark left-handed neutrinos $\nu'^{}_L$.}
\end{figure}

In this sector, we will give the completed model with two or more
gauge-singlet Dirac fermions. The Lagrangian involving the fermion
singlets (FSs) should be
\begin{eqnarray}
\mathcal{L}_{}^{\textrm{M}}\equiv\mathcal{L}_{}^{\textrm{FS}}=\mathcal{L}_K^{\textrm{FS}}
+\mathcal{L}_Y^{\textrm{FS}}+\mathcal{L}_m^{\textrm{FS}}\,,
\end{eqnarray}
with the kinetic terms:
\begin{eqnarray}
\mathcal{L}_K^{\textrm{FS}}=i\bar{N}^{}_{R_i^{}}\! \not\!\partial
N^{}_{R_i^{}} + i\bar{N}'^{}_{R_i^{}}\! \not\!\partial
N'^{}_{R_i^{}}\,,
\end{eqnarray}
the Yukawa terms:
\begin{eqnarray}
\mathcal{L}_Y^{\textrm{FS}}=-(y_N^{})_{ij}^{}\bar{l}^{}_{L_i^{}}
\phi N^{}_{R_j^{}} - (y^{}_{N'})_{ij}^{}\bar{l}'^{}_{L_i^{}} \phi'
N'^{}_{R_j^{}}+\textrm{H.c.}\,,
\end{eqnarray}
and the mass terms:
\begin{eqnarray}
\mathcal{L}_m^{\textrm{FS}}=-(M_{N}^{})_{ij}^{}
\bar{N}'^{c}_{R_i^{}} N^{}_{R_j^{}}+\textrm{H.c.}\,.
\end{eqnarray}
Here we have introduced two types of gauge-singlet right-handed
fermions:
\begin{eqnarray}
\left\{\begin{array}{ll}
N^{}_{R}(\textbf{1},\textbf{1},0)(\textbf{1},\textbf{1},0)&~\textrm{with~a~lepton number}~~+1\,,\\
[2mm]
N'^{}_{R}(\textbf{1},\textbf{1},0)(\textbf{1},\textbf{1},0)&~\textrm{with~a~lepton
number}~~-1\,,
\end{array}\right.
\end{eqnarray}
where the first and second parentheses being the quantum numbers
under the ordinary and dark gauge groups, respectively. Note that
other gauge-invariant Yukawa and mass terms have been forbidden as a
result of the lepton number conservation. After choosing a proper
base, the fermion singlets can have a diagonal and real mass matrix:
\begin{eqnarray}
M_{N}^{}=\textrm{diag}\{M_{N_1^{}}^{}\,,~M_{N_2^{}}^{}\,,...\}\,.
\end{eqnarray}
We then can define the Dirac fermions:
\begin{eqnarray}
N^{}_i=N^{}_{R_i^{}}+N'^{c}_{R_i^{}}\,.
\end{eqnarray}

\subsection{Dirac neutrino masses}

After the $[SU(2)^{}_L]$-doublet Higgs scalar $\phi$ and the
$[SU(2)'^{}_L]$-doublet Higgs scalar $\phi'$ develop their VEVs, the
ordinary and dark left-handed neutrinos as well as the gauge-singlet
right-handed fermions will have a mass matrix as below,
\begin{eqnarray}
\mathcal{L}\supset-\left[\begin{array}{cc}\bar{\nu}^{}_L&\bar{N}'^{c}_R\end{array}\right]
\left[\begin{array}{cc}0& y^{}_N \langle\phi\rangle\\
[2mm]y^T_{N'}  \langle\phi'\rangle & M^{}_N\end{array}\right]
\left[\begin{array}{c}\nu'^{c}_L\\
[2mm] N^{}_R\end{array}\right]+\textrm{H.c.}\,.
\end{eqnarray}
We can block diagonalize the above mass matrix to be
\cite{flv1992,abs1992,berezhiani2005,bb2006-1}
\begin{eqnarray}
\label{idirac}
\mathcal{L}&\supset&-\left[\begin{array}{cc}\bar{\nu}^{}_L&\bar{N}'^{c}_R\end{array}\right]
\left[\begin{array}{cc}m^{}_\nu & 0\\
[2mm]0& M^{}_N\end{array}\right]
\left[\begin{array}{c}\nu'^{c}_L\\
[2mm]
N^{}_R\end{array}\right]+\textrm{H.c.}~~\textrm{with}\nonumber\\
&&m^{}_\nu=-
y^{}_N\frac{\langle\phi\rangle\langle\phi'\rangle}{M^{}_N}y^T_{N'}
\equiv m^{\textrm{I}}_\nu\,,
\end{eqnarray}
if the seesaw condition is satisfied, i.e.
\begin{eqnarray}
y^{}_N \langle\phi\rangle\,,~y^{}_{N'}  \langle\phi'\rangle\ll
M^{}_N\,.
\end{eqnarray}
This means the ordinary and dark left-handed neutrinos will form the
light Dirac neutrinos $\nu=\nu^{}_L + \nu'^{c}_L$, while the
gauge-singlet right-handed fermions will form the heavy Dirac
fermions $N=N^{}_R+N'^{c}_R$. Note that the Dirac neutrino mass
matrix (\ref{idirac}) will have one nonzero eigenvalue if there is
only one gauge-singlet Dirac fermion. We thus need two or more
gauge-singlet Dirac fermions to explain the neutrino oscillation
data. Analogous to the usual type-I seesaw \cite{minkowski1977}
formula of the Majorana neutrino masses, we shall refer to the
formula (\ref{idirac}) of the Dirac neutrino masses to be the type-I
Dirac seesaw. The relevant diagram is shown in Fig. \ref{DiracI}.

\begin{figure*}
\vspace{7.0cm} \epsfig{file=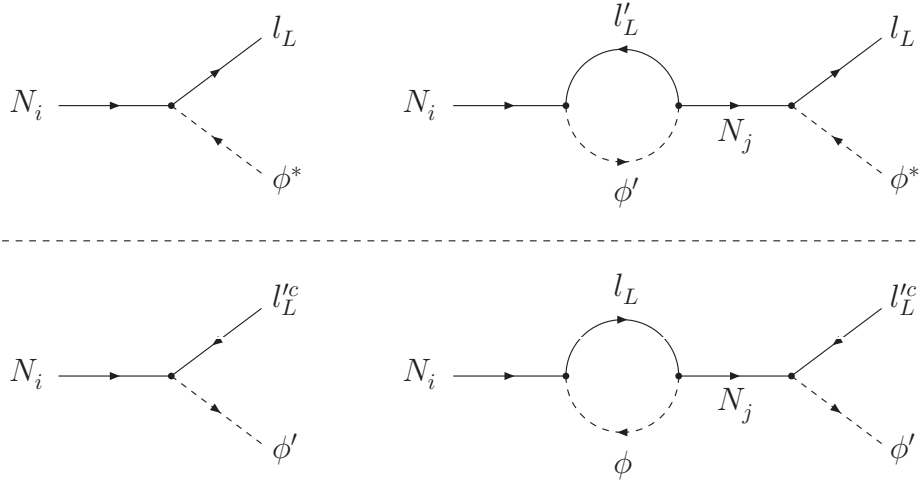, bbllx=5.5cm, bblly=6.0cm,
bburx=15.5cm, bbury=16cm, width=8.5cm, height=8.5cm, angle=0,
clip=0} \vspace{-8.3cm} \caption{\label{Ndecay} The lepton number
conserving decays of the heavy gauge-singlet Dirac fermions
$N_i^{}=N^{}_{R_i^{}}+N'^{c}_{R_i^{}}$ in the type-I Dirac seesaw
scenario. We need at least two fermion singlets to generate a
nonzero lepton asymmetry in the $[SU(2)_L^{}]$-doublet leptons
$l^{}_L$ and an opposite lepton asymmetry in the
$[SU(2)'^{}_L]$-doublet leptons $l'^{}_L$. The CP conjugation is not
shown for simplicity.}
\end{figure*}

\subsection{CP asymmetry}

As shown in Fig. \ref{Ndecay}, the lepton number conserving decays
of the gauge-singlet Dirac fermions $N^{}_i$ can simultaneously
generate a lepton asymmetry $\eta_{l^{}_L}^{}$ in the
$[SU(2)^{}_L]$-doublet leptons $l^{}_L$ and an opposite lepton
asymmetry $\eta_{l'^{}_L}^{}$ in the $[SU(2)'^{}_L]$-doublet leptons
$l'^{}_L$ \footnote{In Ref. \cite{bb2001}, the authors proposed a
novel scenario where the gauge-singlet fermions with heavy Majorana
masses can mediate some lepton number violating scattering processes
that convert the ordinary particles into the dark particles to
produce a lepton asymmetry in the ordinary leptons and an opposite
lepton asymmetry in the dark leptons. This mechanism cannot apply to
the present lepton number conserving model because the cross
sections obey the relations $\sigma(l^{}_L\phi^\ast_{}\rightarrow
l^{}_L\phi^\ast_{})+\sigma(l^{}_L\phi^\ast_{}\rightarrow
l'^{c}_L\phi')=\sigma(l^{c}_L\phi\rightarrow
l^{c}_L\phi)+\sigma(l^{c}_L\phi\rightarrow l'^{}_L\phi'^\ast_{})$,
$\sigma(l^{}_L\phi^\ast_{}\rightarrow
l^{}_L\phi^\ast_{})=\sigma(l^{c}_L\phi\rightarrow l^{c}_L\phi)$ and
then $\sigma(l^{}_L\phi^\ast_{}\rightarrow
l^{}_L\phi^\ast_{})-\sigma(l^{c}_L\phi\rightarrow l^{c}_L\phi)=0$,
$\sigma(l^{}_L\phi^\ast_{}\rightarrow
l'^{c}_L\phi')-\sigma(l^{c}_L\phi\rightarrow
l'^{}_L\phi'^\ast_{})=0$ as a result of CPT invariance.}, i.e.
\begin{eqnarray}
\label{odl}
\eta_{l_L^{}}^{}=-\eta_{l'^{}_L}^{}\propto\varepsilon_{N_i^{}}^{}\,.
\end{eqnarray}
Here the CP asymmetry $\varepsilon_{N_i^{}}^{}$ can be calculated by
\begin{eqnarray}
\label{cpaI}
\varepsilon_{N_i^{}}^{}&=&\frac{\Gamma_{N_i^{}\rightarrow
l_L^{}\phi^\ast_{}}^{}-\Gamma_{N_i^{c}\rightarrow
l_L^{c}\phi}^{}}{\Gamma_{N_i^{}}^{}}=\frac{\Gamma_{N_i^{c}\rightarrow
l'^{}_{L}\phi'^\ast_{}}^{}-\Gamma_{N_i^{}\rightarrow
l'^{c}_{L}\phi'^{}_{}}^{}}{\Gamma_{N_i^{}}^{}}\nonumber\\
&=&-\frac{1}{4\pi}\sum_{j\neq
i}^{}\frac{\textrm{Im}\left[(y_N^\dagger
y_N^{})_{ij}^{}(y^\dagger_{N'}
y^{}_{N'})_{ij}^{}\right]}{(y_N^\dagger
y_N^{})_{ii}^{}+(y^\dagger_{N'}
y^{}_{N'})_{ii}^{}}\frac{M_{N_i^{}}^{}M_{N_j^{}}^{}}{M_{N_j^{}}^2-M_{N_i^{}}^2}\,,\nonumber\\
&&
\end{eqnarray}
with $\Gamma_{N_i^{}}^{}$ being the decay width:
\begin{eqnarray}
\Gamma_{N_i^{}}^{}&=&\Gamma_{N_i^{}\rightarrow
l_L^{}\phi^\ast}^{}+\Gamma_{N_i^{}\rightarrow
l'^{c}_L\phi'}^{}=\Gamma_{N_i^{c}\rightarrow
l^c_L\phi}^{}+\Gamma_{N_i^{c}\rightarrow
l'^{}_L\phi'^\ast_{}}^{}\nonumber\\
&=&\frac{1}{16\pi}[(y_N^\dagger y_N^{})_{ii}^{}+(y^\dagger_{N'}
y^{}_{N'})_{ii}^{}]M_{N_i^{}}^{}\,.
\end{eqnarray}
We should keep in mind that at least two gauge-singlet Dirac
fermions are necessary to induce a nonzero CP asymmetry.

If the gauge-singlet Dirac fermions have a hierarchical mass
spectrum, i.e. $M_{N_i^{}}^2 \ll M_{N_j^{}}^2$, we can simplify the
CP asymmetry (\ref{cpaI}) to be
\begin{eqnarray}
\varepsilon_{N_i^{}}^{} &\simeq&
\frac{1}{4\pi}\frac{\textrm{Im}\left[(y_N^\dagger m^{}_\nu
y^\ast_{N'})_{ii}^{}\right]}{(y_N^\dagger
y_N^{})_{ii}^{}+(y_{N'}^{\dagger}
y^{}_{N'})_{ii}^{}}\frac{M_{N_i^{}}^{}}{\langle\phi\rangle\langle\phi'\rangle}\,.
\end{eqnarray}
Similar to the Davidson-Ibarra bound \cite{di2002} in the type-I
Majorana seesaw scenario, the above CP asymmetry should have an
upper bound:
\begin{eqnarray}
\label{boundI} |\varepsilon_{N_i^{}}^{}| &\leq&
\frac{1}{8\pi}\frac{\left|\textrm{Im}\left[(y_N^\dagger m^{}_\nu
y^\ast_{N'})_{ii}^{}\right]\right|}{\sqrt{(y_N^\dagger
y_N^{})_{ii}^{}(y_{N'}^{\dagger}
y^{}_{N'})_{ii}^{}}}\frac{M_{N_i^{}}^{}}{\langle\phi\rangle\langle\phi'\rangle}\nonumber\\
&<&\frac{1}{8\pi}\frac{M_{N_i^{}}^{}m_{\nu}^{\textrm{max}}}{\langle\phi\rangle\langle\phi'\rangle}
=\varepsilon_{N_i^{}}^{\textrm{max}}\,.
\end{eqnarray}
Here $m_{\nu}^{\textrm{max}}$ is the maximal eigenvalue of the
neutrino mass matrix $m_{\nu}^{}$. Alternatively, the CP asymmetry
(\ref{cpaI}) can be resonantly enhanced \cite{fps1995} if the
gauge-singlet Dirac fermions have a quasi-degenerate mass spectrum,
i.e. $M_{N_i^{}}^2 \simeq M_{N_j^{}}^2 \gg |M_{N_i^{}}^2 -
M_{N_j^{}}^2|$.

\section{The model with $[SU(2)_L^{}\times SU(2)'^{}_L]$-bidoublet Higgs
scalars}

In this sector, we will give the completed model with two or more
$[SU(2)_L^{}\times SU(2)'^{}_L]$-bidoublet Higgs scalars:
\begin{eqnarray}
\Sigma_{a}^{}(\textbf{1},\textbf{2},-1)(\textbf{1},\textbf{2},+1)
=\left[\begin{array}{ll}\bar{\sigma}^{+}_{a}&\sigma^{0}_{a}\\
[1mm] \bar{\sigma}^{0}_{a}&\sigma^{-}_{a}\end{array}\right]\,.
\end{eqnarray}
Here the first and second parentheses stand for the quantum numbers
under the ordinary and dark gauge groups, respectively. The
Lagrangian involving the Higgs bidoublets (HBs) should be
\begin{eqnarray}
\mathcal{L}_{}^{\textrm{M}}\equiv\mathcal{L}_{}^{\textrm{HB}}=\mathcal{L}_K^{\textrm{HB}}
+\mathcal{L}_Y^{\textrm{HB}}-V^{\textrm{HB}}_{}\,,
\end{eqnarray}
where the kinetic terms are
\begin{eqnarray}
\mathcal{L}_K^{\textrm{HB}}&=&\textrm{Tr}[(D_\mu^{}\Sigma^{}_a)^\dagger_{}(D^\mu_{}\Sigma^{}_a)]~~\textrm{with}\nonumber\\
&& D_\mu^{}\Sigma^{}_a=\partial_\mu^{}\Sigma^{}_a+i
\frac{1}{2}g_1^{} B_\mu^{}\Sigma^{}_a-i g_2^{}
\frac{\tau_i^{}}{2}W^i_{\mu}\Sigma^{}_a \nonumber\\
&&\quad\quad\quad\quad-i \frac{1}{2}g'^{}_1 B'^{}_\mu\Sigma^{}_a - i
g'^{}_2 \Sigma^{}_a\frac{\tau_i^{T}}{2}W'^i_{\mu}\,,
\end{eqnarray}
the Yukawa couplings contain
\begin{eqnarray}
\mathcal{L}_Y^{\textrm{HB}}&=&-f_{a}^{}\bar{l}^{}_L \Sigma_a^{}
l'^{c}_L+\textrm{H.c.}\,,
\end{eqnarray}
and the scalar potential includes
\begin{eqnarray}
V^{\textrm{HB}}_{}&=&(M_{\Sigma}^2)_{ab}^{}\textrm{Tr}(\Sigma^\dagger_{a}\Sigma^{}_b)
+(\lambda_\Sigma^{})_{abcd}^{}\textrm{Tr}(\Sigma^\dagger_{a}\Sigma_b^{})
\textrm{Tr}(\Sigma^\dagger_{c}\Sigma_d^{})\nonumber\\
&& +2[(\lambda_{\phi\Sigma}^{})_{ab}^{}\phi^\dagger_{}\phi
+(\lambda_{\phi'\Sigma}^{})_{ab}^{}\phi'^\dagger_{}\phi']
\textrm{Tr}(\Sigma^\dagger_{a}\Sigma_b^{})\nonumber\\
&& +2[(\lambda_{\delta\Sigma}^{})_{ab}^{}\delta^\dagger_{}\delta
+(\lambda_{\delta'\Sigma}^{})_{ab}^{}\delta'^\dagger_{}\delta']
\textrm{Tr}(\Sigma^\dagger_{a}\Sigma_b^{})\nonumber\\
&&+\rho_a^{}\phi^\dagger \Sigma_a^{} \phi'^\ast_{} +
\textrm{H.c.}\,.
\end{eqnarray}
Without loss of generality, we can choose a base to take
\begin{eqnarray}
M_\Sigma^{2}=\textrm{diag}\{M_{\Sigma_1^{}}^{2}\,,~M_{\Sigma_2^{}}^{2}\,,~...\}\,,
~\rho_1^{}=\rho_1^\ast\,,~\rho_2^{}=\rho_2^\ast\,,...\,.
\end{eqnarray}
Note that the above Yukawa couplings and scalar potential will
exactly conserve the lepton number because the Higgs bidoublets
$\Sigma_{a}^{}$ don't carry any lepton numbers.

\subsection{Dirac neutrino masses}

\begin{figure}
\vspace{6cm} \epsfig{file=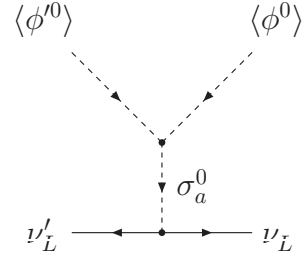, bbllx=3cm, bblly=6.0cm,
bburx=13cm, bbury=16cm, width=8.5cm, height=8.5cm, angle=0, clip=0}
\vspace{-10cm} \caption{\label{DiracII} The type-II Dirac seesaw for
generating the masses between the ordinary left-handed neutrinos
$\nu_L^{}$ and the dark left-handed neutrinos $\nu'^{}_L$.}
\end{figure}

After the $[SU(2)^{}_L]$-doublet Higgs scalar $\phi$ and the
$[SU(2)'^{}_L]$-doublet Higgs scalar $\phi'$ acquire their VEVs, the
heavy $[SU(2)_L^{}\times SU(2)'^{}_L]$-bidoublet Higgs scalars
$\Sigma^{}_a$ can pick up the seesaw-suppressed VEVs:
\begin{eqnarray}
\langle\Sigma_{a}^{}\rangle
\!\!&=&\!\!\left[\begin{array}{cc}0&\langle\sigma^{0}_a\rangle\\
[1mm]
0&0\end{array}\right]\,\textrm{with}\,\langle\sigma^{0}_a\rangle\simeq
-\frac{\rho^{}_a
\langle\phi\rangle\langle\phi'\rangle}{M^{2}_{\Sigma_a^{}}}\ll\langle\phi\rangle\,,\langle\phi'\rangle\,.\nonumber\\
&&
\end{eqnarray}
We hence can naturally obtain the light Dirac neutrinos composed of
the ordinary left-handed neutrinos $\nu^{}_L$ and the dark
left-handed neutrinos $\nu'^{}_L$, i.e.
\begin{eqnarray}
\label{iidirac} \mathcal{L}&\supset&-m^{}_\nu \bar{\nu}^{}_L
\nu'^c_L+\textrm{H.c.}~~\textrm{with}\nonumber\\
&&m^{}_\nu=\sum_a^{}f^{}_a \langle\Sigma^{}_a\rangle\equiv
\sum_a^{}m^{\textrm{II}a}_{\nu}\equiv m^{\textrm{II}}_{\nu}\,.
\end{eqnarray}
The above mechanism for the Dirac neutrino masses is very similar to
the usual type-II seesaw \cite{mw1980} for the Majorana neutrino
masses. So, it may be named as the type-II Dirac seesaw. We show the
relevant diagram in Fig. \ref{DiracII}.

\subsection{CP asymmetry}

\begin{figure*}
\vspace{7.0cm} \epsfig{file=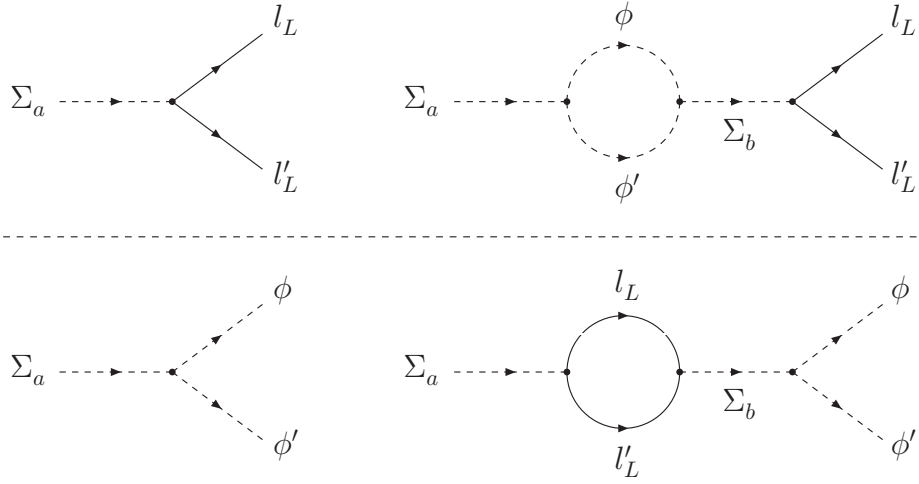, bbllx=5.5cm, bblly=6.0cm,
bburx=15.5cm, bbury=16cm, width=8.5cm, height=8.5cm, angle=0,
clip=0} \vspace{-8.3cm} \caption{\label{Sdecay} The lepton number
conserving decays of the heavy $[SU(2)_L^{}\times
SU(2)'^{}_L]$-bidoublet Higgs scalars $\Sigma^{}_a$ in the type-II
Dirac seesaw scenario. We need at least two Higgs bidoublets to
generate a nonzero lepton asymmetry in the $[SU(2)_L^{}]$-doublet
leptons $l^{}_L$ and an opposite lepton asymmetry in the
$[SU(2)'^{}_L]$-doublet leptons $l'^{}_L$. The CP conjugation is not
shown for simplicity.}
\end{figure*}

From Fig. \ref{Sdecay}, it is straightforward to see that the lepton
number conserving decays of the $[SU(2)_L^{}\times
SU(2)'^{}_L]$-bidoublet Higgs scalars $\Sigma_{a}^{}$ can generate a
lepton asymmetry $\eta_{l^{}_L}^{}$ in the $[SU(2)_L^{}]$-doublet
leptons $l_L^{}$ and an opposite lepton asymmetry
$\eta_{l'^{}_L}^{}$ in the $[SU(2)'^{}_L]$-doublet leptons
$l'^{}_L$, i.e.
\begin{eqnarray}
\eta_{l_L^{}}^{}=-\eta_{l'^{}_L}^{}\propto\varepsilon_{\Sigma_a^{}}^{}\,,
\end{eqnarray}
where $\varepsilon_{\Sigma_a^{}}^{}$ is the CP asymmetry:
\begin{eqnarray}
\varepsilon_{\Sigma_{a}^{}}^{}&=&\frac{\Gamma_{\Sigma_a^{}\rightarrow
l_L^{}l'^{}_L}^{}-\Gamma_{\Sigma^\ast_{a}\rightarrow
l_L^{c}l'^{c}_L}^{}}{\Gamma_{\Sigma_a^{}}^{}}\nonumber\\
&=&\frac{\Gamma_{\Sigma^\ast_{a}\rightarrow
\phi^\ast_{}\phi'^\ast_{}}^{}-\Gamma_{\Sigma^{}_a\rightarrow
\phi\phi'}^{}}{\Gamma_{\Sigma_a^{}}^{}}\,,
\end{eqnarray}
with $\Gamma_{\Sigma_a^{}}^{}$ being the decay width:
\begin{eqnarray}
\Gamma_{\Sigma_a^{}}^{}&=&\Gamma_{\Sigma_a^{}\rightarrow
l_L^{}l'^{}_L}^{}+\Gamma_{\Sigma_a^{}\rightarrow
\phi\phi'}^{}\nonumber\\
&=&\Gamma_{\Sigma_a^\ast\rightarrow l_L^c
l'^{c}_{L}}^{}+\Gamma_{\Sigma_a^\ast\rightarrow
\phi^\ast_{}\phi'^\ast_{}}^{}\,.
\end{eqnarray}
We can calculate the decay width at tree level,
\begin{eqnarray}
\Gamma_{\Sigma_a^{}}^{}=\frac{1}{16\pi}\left[\textrm{Tr}\left(f_a^\dagger
f_a^{}\right)+\frac{\rho_a^{2}}{M_{\Sigma_a^{}}^{2}}\right]M_{\Sigma_a^{}}^{}\,,
\end{eqnarray}
and the CP asymmetry at one-loop order,
\begin{eqnarray}
\label{cpaII}
\varepsilon_{\Sigma_a^{}}^{}=-\frac{1}{4\pi}\sum_{b\neq
a}^{}\frac{\textrm{Im}\left[\textrm{Tr}\left(f_b^\dagger
f_a^{}\right)\right]}{\textrm{Tr}\left(f_a^\dagger
f_a^{}\right)+\frac{\rho_a^{2}}{M_{\Sigma_a^{}}^{2}}}
\frac{\rho_b^{}\rho_a^{}}{M_{\Sigma_b^{}}^2-M_{\Sigma_a^{}}^2}\,.
\end{eqnarray}
Note that at least two $[SU(2)_L^{}\times SU(2)'^{}_L]$-bidoublet
Higgs scalars should be introduced to generate a nonzero CP
asymmetry.

Obviously, the above CP asymmetry can be resonantly enhanced if the
$[SU(2)_L^{}\times SU(2)'^{}_L]$-bidoublet Higgs scalars have a
quasi-degenerate mass spectrum, i.e. $M_{\Sigma_a^{}}^2 \simeq
M_{\Sigma_b^{}}^2 \gg |M_{\Sigma_a^{}}^2 - M_{\Sigma_b^{}}^2|$. In
the hierarchical case with $M_{\Sigma_a^{}}^2\ll M_{\Sigma_b^{}}^2$,
the CP asymmetry (\ref{cpaII}) can be simplified by
\begin{eqnarray}
\varepsilon_{\Sigma_a^{}}^{}
\simeq-\frac{1}{4\pi}\frac{\textrm{Im}\left\{\textrm{Tr}\left[\left(\sum_{b\neq
a}^{}m^{\textrm{II}b\dagger}_{\nu}\right)
m^{\textrm{II}a}_{\nu}\right]\right\}}{\textrm{Tr}\left(f_a^\dagger
f_a^{}\right)+\frac{\rho_a^{2}}{M_{\Sigma_a^{}}^{2}}}
\frac{M_{\Sigma_a^{}}^2}{\langle\phi\rangle^2_{}\langle\phi'\rangle^2_{}}\,,
\end{eqnarray}
and then have an upper bound:
\begin{eqnarray}
|\varepsilon_{\Sigma_a^{}}^{}|
&\leq&\frac{1}{8\pi}\frac{\left|\textrm{Im}\left\{\textrm{Tr}\left[\left(\sum_{b\neq
a}^{}m^{\textrm{II}b\dagger}_{\nu}\right)
m^{\textrm{II}a}_{\nu}\right]\right\}\right|}{\sqrt{\textrm{Tr}\left(f_a^\dagger
f_a^{}\right)\frac{\rho_a^{2}}{M_{\Sigma_a^{}}^{2}}}}
\frac{M_{\Sigma_a^{}}^2}{\langle\phi\rangle^2_{}\langle\phi'\rangle^2_{}}\nonumber\\
&=&\frac{1}{8\pi}\frac{\left|\textrm{Im}\left\{\textrm{Tr}\left[\left(\sum_{b\neq
a}^{}m^{\textrm{II}b\dagger}_{\nu}\right)
m^{\textrm{II}a}_{\nu}\right]\right\}\right|}{\sqrt{\textrm{Tr}\left(m^{\textrm{II}a\dagger}_{\nu}
m^{\textrm{II}a}_{\nu}\right)}}
\frac{M_{\Sigma_a^{}}^{}}{\langle\phi\rangle\langle\phi'\rangle}\nonumber\\
&<&\frac{1}{8\pi}
\frac{M_{\Sigma_a^{}}^{}m^{\textrm{II}b\textrm{max}}_{\nu}}{\langle\phi\rangle\langle\phi'\rangle}
=\varepsilon_{\Sigma_a^{}}^{\textrm{max}}\,.
\end{eqnarray}
Here $m_{\nu}^{\textrm{II}b\textrm{max}}$ is the maximal eigenvalue
of the mass matrix $\sum_{b\neq a}^{}m_{\nu}^{\textrm{II}b}$. Unless
there is a large cancellation between the mass matrix
$m_{\nu}^{\textrm{II}a}$ and the mass matrix $\sum_{b\neq
a}^{}m_{\nu}^{\textrm{II}b}$, the eigenvalue
$m_{\nu}^{\textrm{II}b\textrm{max}}$ will not be much bigger than
the largest neutrino mass $m_{\nu}^{\textrm{max}}$. So, we can
roughly constrain
\begin{eqnarray}
\label{boundII} |\varepsilon_{\Sigma_a^{}}^{}| &<&\frac{1}{8\pi}
\frac{M_{\Sigma_a^{}}^{}m^{\textrm{max}}_{\nu}}{\langle\phi\rangle\langle\phi'\rangle}\,.
\end{eqnarray}

\section{The model with gauge-singlet Dirac fermion(s) and $[SU(2)_L^{}\times SU(2)'^{}_L]$-bidoublet Higgs
scalar(s)}

In this sector, we will give the completed model with at least one
gauge-singlet Dirac fermion and at least one $[SU(2)_L^{}\times
SU(2)'^{}_L]$-bidoublet Higgs scalar. By taking the notations in the
previous sections, the Lagrangian involving the fermion singlet(s)
and the Higgs bidoublet(s) can be described by
\begin{eqnarray}
\mathcal{L}_{}^{\textrm{M}}\equiv\mathcal{L}_{}^{\textrm{FS}}+\mathcal{L}_{}^{\textrm{HB}}\,.
\end{eqnarray}

\subsection{Dirac neutrino masses}

The ordinary and dark left-handed neutrinos as well as the
gauge-singlet right-handed fermions should have the following mass
matrix:
\begin{eqnarray}
\mathcal{L}&\supset&-\left[\begin{array}{cc}\bar{\nu}^{}_L&\bar{N}'^{c}_R\end{array}\right]
\left[\begin{array}{cc}\sum_a^{}f^{}_a \langle\Sigma^{}_a\rangle & y^{}_N \langle\phi\rangle\\
[2mm]y^T_{N'}  \langle\phi'\rangle & M^{}_N\end{array}\right]
\left[\begin{array}{c}\nu'^{c}_L\\
[2mm] N^{}_R\end{array}\right]\nonumber\\
&&+\textrm{H.c.}\,.
\end{eqnarray}
Obviously, the ordinary left-handed neutrinos $\nu^{}_L$ and the
dark left-handed neutrinos $\nu'^{}_L$ will form three light Dirac
neutrinos as their mass term is just a sum of the type-I Dirac
seesaw (\ref{idirac}) and the type-II Dirac seesaw (\ref{iidirac}),
i.e.
\begin{eqnarray}
\label{ipiidirac} m^{}_\nu= m^{\textrm{I}}_\nu +
m^{\textrm{II}}_\nu\,.
\end{eqnarray}

\begin{figure*}
\vspace{7.0cm} \epsfig{file=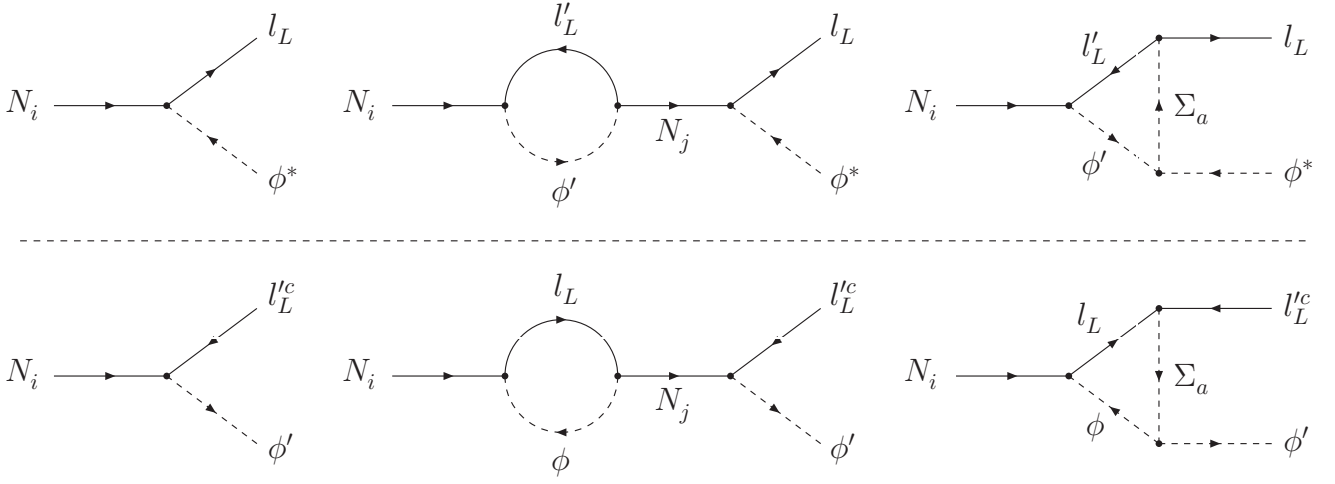, bbllx=5.5cm, bblly=6.0cm,
bburx=15.5cm, bbury=16cm, width=8.5cm, height=8.5cm, angle=0,
clip=0} \vspace{-8.3cm} \caption{\label{NSdecay} The lepton number
conserving decays of the heavy gauge-singlet Dirac fermions
$N_i^{}=N^{}_{R_i^{}}+N'^{c}_{R_i^{}}$ in the type-I+II Dirac seesaw
scenario. If there is only one fermion singlet, the self-energy
correction will not contribute to the generation of the lepton
asymmetries in the $[SU(2)_L^{}]$-doublet leptons $l^{}_L$ and the
$[SU(2)'^{}_L]$-doublet leptons $l'^{}_L$. The CP conjugation is not
shown for simplicity.}
\end{figure*}

\subsection{CP asymmetry}

\begin{figure*}
\vspace{7.0cm} \epsfig{file=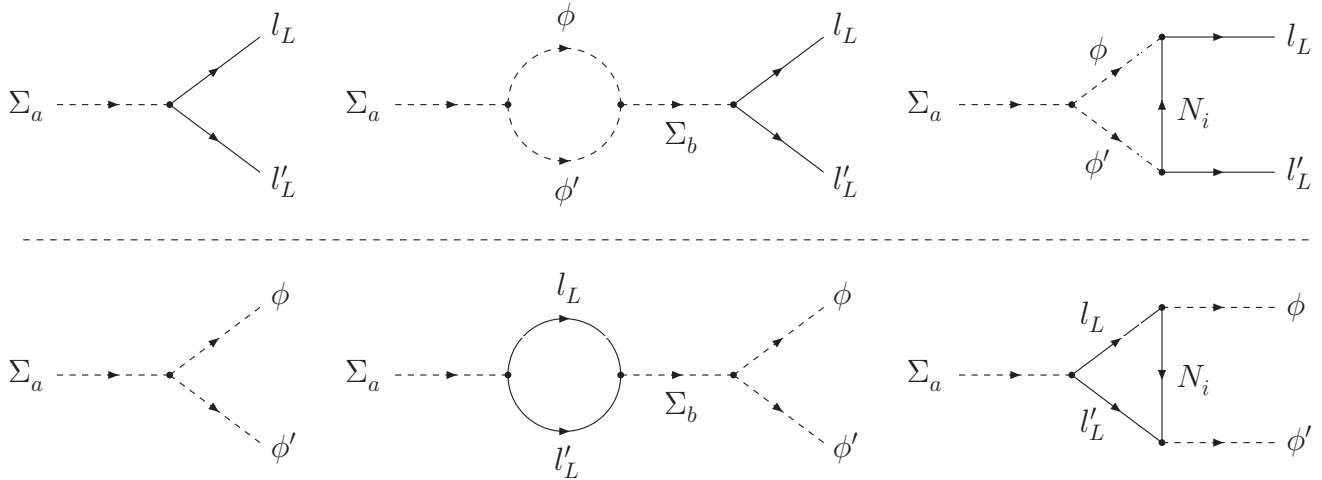, bbllx=5.5cm, bblly=6.0cm,
bburx=15.5cm, bbury=16cm, width=8.5cm, height=8.5cm, angle=0,
clip=0} \vspace{-8.3cm} \caption{\label{SNdecay} The lepton number
conserving decays of the heavy $[SU(2)_L^{}\times
SU(2)'^{}_L]$-bidoublet Higgs scalars $\Sigma^{}_a$ in the type-I+II
Dirac seesaw scenario. If there is only one Higgs bidoublet, the
self-energy correction will not contribute to the generation of the
lepton asymmetries in the $[SU(2)_L^{}]$-doublet leptons $l^{}_L$
and the $[SU(2)'^{}_L]$-doublet leptons $l'^{}_L$. The CP
conjugation is not shown for simplicity.}
\end{figure*}

As shown in Figs. \ref{NSdecay} and \ref{SNdecay}, the lepton number
conserving decays of the heavy gauge-singlet Dirac fermions $N_i^{}$
and/or $[SU(2)_L^{}\times SU(2)'^{}_L]$-bidoublet Higgs scalars
$\Sigma^{}_a$ can simultaneously generate a lepton asymmetry in the
$[SU(2)^{}_L]$-doublet leptons $l^{}_L$ and an opposite lepton
asymmetry in the $[SU(2)'^{}_L]$-doublet leptons $l'^{}_L$. The
relevant CP asymmetries should be
\begin{eqnarray}
\label{cpaNS} \varepsilon_{N_i^{}}^{}
&=&-\frac{1}{4\pi}\frac{1}{(y_N^\dagger
y_N^{})_{ii}^{}+(y^\dagger_{N'} y^{}_{N'})_{ii}^{}}\nonumber\\
&&\times \left\{\sum_{j\neq i}^{}\textrm{Im}\left[(y_N^\dagger
y_N^{})_{ij}^{}(y^\dagger_{N'}
y^{}_{N'})_{ij}^{}\right]\frac{M_{N_i^{}}^{}M_{N_j^{}}^{}}{M_{N_j^{}}^2-M_{N_i^{}}^2}\right.\nonumber\\
&& +2\,\textrm{Im}\left[(y^\dagger_N f^{}_a
y^\ast_{N'})_{ii}^{}\right]\frac{\rho^{}_a}{M_{N_i^{}}^{}}\nonumber\\
&&\times \left.\left[1-\frac{M_{\Sigma_a^{}}^2}{M_{N_i^{}}^2}
\ln\left(1+\frac{M_{N_i^{}}^2}{M_{\Sigma_a^{}}^2}\right)\right]\right\}\,,
\end{eqnarray}
and
\begin{eqnarray}
\label{cpaSN} \varepsilon_{\Sigma_a^{}}^{}
&=&-\frac{1}{4\pi}\frac{1}{\textrm{Tr}(f^\dagger_a
f^{}_a)+\frac{\rho_a^2}{M_{\Sigma_a^{}}^2}}\left\{\sum_{b\neq a}^{}
\frac{\textrm{Im}\left[\textrm{Tr}\left(f_b^\dagger
f_a^{}\right)\right]\rho_b^{}\rho_a^{}}{M_{\Sigma_b^{}}^2-M_{\Sigma_a^{}}^2}\right.\nonumber\\
&& \left.+\sum_{i}^{}\textrm{Im}\left[(y^\dagger_N f^{}_a
y^{\ast}_{N'})_{ii}^{}\right]\frac{\rho_a^{}M_{N_i^{}}^{}}{M_{\Sigma_a^{}}^{2}}\ln
\left(1+\frac{M_{\Sigma_a^{}}^2}{M_{N_i^{}}^2}\right)\right\}\,.\nonumber\\
&&
\end{eqnarray}
Clearly, the vertex correction should be the unique source for the
nonzero CP asymmetry (\ref{cpaNS}) if there is only one
gauge-singlet Dirac fermion. As for the CP asymmetry (\ref{cpaSN}),
it will not be affected by the self-energy correction if we only
introduce one $[SU(2)_L^{}\times SU(2)'^{}_L]$-bidoublet Higgs
scalar.

Similar to those in the pure type-I and type-II Dirac seesaw models,
the CP asymmetries (\ref{cpaNS}) and (\ref{cpaSN}) can be simplified
in the hierarchical cases, i.e.
\begin{eqnarray}
\varepsilon_{N_i^{}}^{} &\simeq&
\frac{1}{4\pi}\frac{\textrm{Im}\left[(y_N^\dagger m^{}_\nu
y^\ast_{N'})_{ii}^{}\right]}{(y_N^\dagger
y_N^{})_{ii}^{}+(y^\dagger_{N'}
y^{}_{N'})_{ii}^{}}\frac{M_{N_i^{}}^{}}{\langle\phi\rangle\langle\phi'\rangle}\nonumber\\
&&~~\textrm{for}~~M_{N_i^{}}^{2}\ll
M_{N_j^{}}^{2}\,,M_{\Sigma_a^{}}^{2}\,,\\
\varepsilon_{\Sigma_a^{}}^{}
&\simeq&-\frac{1}{4\pi}\frac{\textrm{Im}\left\{\textrm{Tr}[(m^{\textrm{I}}_\nu
+ \sum_{b\neq a}^{}m^{\textrm{II}b}_\nu)^\dagger_{}
m^{\textrm{II}a}_\nu ]\right\}}{\textrm{Tr}(f^\dagger_a
f^{}_a)+\frac{\rho^{2}_a}{M_{\Sigma_a^{}}^2}}\frac{M_{\Sigma_a^{}}^{2}}
{\langle\phi\rangle^2_{}\langle\phi'\rangle^2_{}}\nonumber\\
&&~~\textrm{for}~~ M_{\Sigma_a^{}}^{2}\ll
M_{\Sigma_b^{}}^{2}\,,~M_{N_i^{}}^{2}\,,
\end{eqnarray}
and have the upper bounds given in Eqs. (\ref{boundI}) and
(\ref{boundII}).

\section{Ordinary and dark baryon asymmetries}

The $SU(2)_L^{}$ sphaleron processes \cite{krs1985} will partially
transfer the lepton asymmetry $\eta_{l^{}_L}^{}$ to a baryon
asymmetry $\eta_B^{}$ in the ordinary sector,
\begin{eqnarray}
\label{obasymmetry} \eta_B^{}=-\frac{28}{79}\eta_{l_L^{}}^{}\,.
\end{eqnarray}
On the other hand, the lepton asymmetry $\eta_{l'^{}_L}^{}$ will
result in a baryon asymmetry $\eta'^{}_{B}$ in the dark sector,
\begin{eqnarray}
\label{dbasymmetry} \eta'^{}_{B}=-\frac{28}{79}\eta_{l'^{}_L}^{}\,,
\end{eqnarray}
through the $SU(2)'^{}_L$ sphaleron processes \footnote{After the
$U(1)'^{}_{em}$ symmetry is broken, there will be a lepton number
violating mass term of the dark right-handed charged leptons, i.e.
the $\delta m_{e'}^{}$ term in Eq. (\ref{dfmass}). We hence require
the $U(1)'^{}_{em}$ symmetry breaking to happen at a lower
temperature
$T_{\textrm{em}'}^{}=\mathcal{O}(\langle\delta'\rangle)<T^{}_{\textrm{sph}'}=\mathcal{O}(\langle\phi'\rangle)$,
where the $SU(2)'^{}_L$ sphaleron processes have become very weak,
to avoid the washout of the dark baryon asymmetry.}. So, the
correlation between the baryon asymmetries in the ordinary and dark
sectors should be
\begin{eqnarray}
\eta'^{}_{B}=-\eta_{B}^{}
~~\textrm{for}~~\eta_{l_L^{}}^{}=-\eta_{l'^{}_L}^{}\,.
\end{eqnarray}
If the dark matter relic density is dominated by the dark proton
and/or neutron, we should have
\begin{eqnarray}
\Omega_B^{}h^2_{}:\Omega_{DM}^{}h^2_{}&=&m_p^{}\eta_B^{}:m_{p'(n')}^{}(-\eta'^{}_{B})
=m_p^{}:m_{p'(n')}^{}\nonumber\\
&\simeq& 1:5\,,
\end{eqnarray}
to fit the cosmological observations. In this case, the dark matter
mass should be determined by \cite{berezhiani2005,berezhiani2006}
\footnote{The sphaleron processes in the ordinary and dark sectors
could work at different temperatures. For example, the sphalerons in
the ordinary sector can keep in equilibrium till a lower temperature
$T^{}_{\textrm{sph}}=\mathcal{O}(\langle\phi\rangle)$ after those in
the dark sector go out of equilibrium at a higher temperature
$T^{}_{\textrm{sph}'}=\mathcal{O}(\langle\phi'\rangle)$. In Eqs.
(\ref{obasymmetry}) and (\ref{dbasymmetry}), we have assumed all of
the lepton asymmetries to be produced during the epoch that the
ordinary and dark sphalerons are both active. Alternatively, the
generation of the lepton asymmetries can cross the critical
temperature $T^{}_{\textrm{sph}'}$. In this case, the lepton
asymmetries generated after this temperature can contribute to the
ordinary baryon asymmetry but cannot affect the dark baryon
asymmetry. This is like the mechanism in the $\nu$MSM model for
generating a large lepton asymmetry but a small baryon asymmetry
\cite{abs2005}. For a proper choice of the masses and couplings of
the decaying gauge-singlet Dirac fermion(s) and/or
$[SU(2)^{}_L\times SU(2)'^{}_L]$-bidoublet Higgs scalar(s), the dark
baryon asymmetry can be smaller than the ordinary baryon asymmetry.
In consequence, the lightest dark nucleon as the dark matter
particle should have a mass bigger than $5\,\textrm{GeV}$.}
\begin{eqnarray}
m_{p'(n')}^{}\simeq 5\,\textrm{GeV}\,.
\end{eqnarray}

After the gauge-singlet Dirac fermions and/or the $[SU(2)^{}_L\times
SU(2)'^{}_L]$-bidoublet Higgs scalars are thermally produced and
then go out of equilibrium, their CP-violating decays can generate
the required ordinary and dark baryon asymmetries. For this purpose,
we can choose the proper values of the masses and couplings of the
decaying particles, like those in the usual leptogenesis within the
Majorana seesaw models \cite{dnn2008}.

\subsection{Baryon asymmetries from the decays of gauge-singlet
Dirac fermions}

In order to show the decays of gauge-singlet Dirac fermions can
generate the required baryon asymmetries, as an example, let us
consider a type-I Dirac seesaw model with two hierarchical fermion
singlets $N_{1,2}^{}$. In this case, the neutrino mass matrix
(\ref{idirac}) is clearly rank 2 so that one neutrino mass
eigenstate should be massless and then the maximal neutrino mass
eigenvalue should be about \cite{ftv2012}
\begin{eqnarray}
m_{\nu}^{\textrm{max}}\simeq 0.05\,\textrm{eV}\,.
\end{eqnarray}
Without loss of generality, we assume $N_1^{}$ much lighter than
$N_{2}^{}$. The final baryon asymmetries then could mainly come from
the decays of the lighter $N_1^{}$. If the Yukawa couplings
$(y_{N}^{})_{i1}^{}$ and $(y_{N'}^{})_{i1}^{}$ of the decaying
$N_1^{}$ give
\begin{eqnarray}
\label{oeqn}
K_{N_1^{}}^{}=\frac{\Gamma_{N_1^{}}^{}}{2H(T)}\left|_{T=M_{N_1^{}}^{}}^{}\right.\ll
1\,,
\end{eqnarray}
the out-of-equilibrium condition is well satisfied and there is no
washout effect associated to the $(y_{N}^{})_{i1}^{}$ and
$(y_{N'}^{})_{i1}^{}$ couplings. As a result, the final baryon
asymmetry can be well described by \cite{kt1990},
\begin{eqnarray}
\eta_B^{}&=&-\frac{28}{79}\eta_{l_L^{}}^{}\simeq
-\frac{28}{79}\times \frac{\varepsilon_{N_1^{}}^{}}{g_\ast^{}}\,.
\end{eqnarray}
Here and thereafter
\begin{eqnarray}
H(T)=\left(\frac{8\pi^{3}_{}g_{\ast}^{}}{90}\right)^{\frac{1}{2}}_{}
\frac{T^{2}_{}}{M_{\textrm{Pl}}^{}}\,,
\end{eqnarray}
is the Hubble constant with $M_{\textrm{Pl}}^{}\simeq 1.22\times
10^{19}_{}\,\textrm{GeV}$ being the Planck mass and
$g_\ast^{}=2\times(106.75+2)=217.5$ being the relativistic degrees
of freedom. For a quantitative estimation, we take
\begin{eqnarray}
\langle\phi'\rangle=500\,\langle\phi\rangle\,,~M_{N_1^{}}^{}
=0.01\,M_{N_{2}^{}}^{}&=&10^{13}_{}\,\textrm{GeV}\,,\nonumber\\
(y_N^{})_{i1}^{}=(y_{N'}^{})_{i1}^{}\sim
0.1(y_N^{})_{i2}^{}&=&0.1(y_{N'}^{})_{i2}^{}\nonumber\\
&=&\mathcal{O}(0.01)\,, \quad~~
\end{eqnarray}
so that the elements of the neutrino mass matrix (\ref{idirac}) can
arrive at the values of the order of
$\mathcal{O}(0.1\,\textrm{eV})$. The above parameter choice also
leads to
\begin{eqnarray}
K_{N_1^{}}^{}=
\mathcal{O}(0.01-0.1)\,,~\varepsilon_{N_1^{}}^{\textrm{max}}\simeq
1.3\times 10^{-5}_{}\,.
\end{eqnarray}
Therefore, we can obtain the needed baryon asymmetry $\eta_B^{}\sim
10^{-10}_{}$ \cite{komatsu2010} as the CP asymmetry
$\varepsilon_{N_1^{}}^{}$ is allowed to be of the order of
$\mathcal{O}(10^{-7}_{})$.

\subsection{Baryon asymmetries from the decays of $[SU(2)_L^{}\times SU(2)'^{}_L]$-bidoublet Higgs scalars}

We now study the baryon asymmetries from the decays of
$[SU(2)_L^{}\times SU(2)'^{}_L]$-bidoublet Higgs scalars. As an
example, let us consider the type-II Dirac seesaw model with two
hierarchical Higgs bidoublets $\Sigma_{1,2}^{}$. We take
\begin{eqnarray}
\label{iip}
&&\langle\phi'\rangle=500\langle\phi\rangle\,,~M_{\Sigma_1^{}}^{}=0.01\,M_{\Sigma_2^{}}^{}=10^{14}_{}\,\textrm{GeV}\,,\nonumber\\
&&\rho_1^{}=0.01\,\rho_2^{}= 2\times 10^{12}_{}\,\textrm{GeV}\,,
\end{eqnarray}
to give
\begin{eqnarray}
\langle\Sigma_1^{}\rangle=100\langle\Sigma_2^{}\rangle\simeq
-3\,\textrm{eV}\,.
\end{eqnarray}
We then assume
\begin{eqnarray}
f_2^{}=f_1^{}e^{i\delta}_{}
\end{eqnarray}
to simplify the neutrino mass matrix as
\begin{eqnarray}
m_\nu^{}=f_1^{}\langle\Sigma_1^{}\rangle+
f_2^{}\langle\Sigma_2^{}\rangle\simeq
f_1^{}\langle\Sigma_1^{}\rangle\,.
\end{eqnarray}
By inputting
\begin{eqnarray}
\textrm{Tr}(f_1^\dagger
f_1^{})&\simeq&\frac{\textrm{Tr}(m_\nu^\dagger
m_\nu^{})}{\langle\Sigma_1^{}\rangle^2_{}}=\frac{\sum_{i}^{}m_i^2}{\langle\Sigma_1^{}\rangle^2_{}}=
1.1\times 10^{-3}_{}\nonumber\\
&&\textrm{for}~~\sum_{i}^{}m_i^2=0.01\,\textrm{eV}^2_{}\,,
\end{eqnarray}
we can read the CP asymmetry
\begin{eqnarray}
\varepsilon_{\Sigma_1^{}}^{}=-2.3\times 10^{-7}_{}\sin\delta\,.
\end{eqnarray}
The above parameter choice also satisfies the out-of-equilibrium
condition in the weak washout region,
\begin{eqnarray}
K=\frac{\Gamma_{\Sigma_1^{}}^{}}{2
H(T)}\left|_{T=M_{\Sigma_1^{}}^{}}^{}\right.\simeq 0.08\,.
\end{eqnarray}
Note for the heavy masses in Eq. (\ref{iip}), the gauge interactions
of the Higgs bidoublets can be safely kept out of equilibrium at the
leptogenesis epoch \cite{hms2000,hrs2005}. The final baryon
asymmetry thus can be approximately calculated by \cite{kt1990},
\begin{eqnarray}
\eta_B^{}&=&-\frac{28}{79}\eta_{l_L^{}}^{}\simeq
-\frac{28}{79}\times
\frac{\varepsilon_{\Sigma_1^{}}^{}}{g_\ast^{}}\times
4\nonumber\\
&=&0.886\times 10^{-10}_{}\left(\frac{\sin\delta}{0.59}\right)\,,
\end{eqnarray}
which is consistent with the measured value \cite{komatsu2010}. Here
we have added the factor $4$ because the decaying particles are the
$[SU(2)_L\times SU(2)_R^{}]$-bidoublet Higgs scalars.

\section{Other constraints and implications}

The dark photon $A'$ can decay into the ordinary fermion pairs
$f\bar{f}$ and the dark fermion pairs $f'\bar{f'}$ as long as the
kinematics is allowed. For example, if the dark photon is much
heavier than the ordinary down quark but lighter than the ordinary
strange quark and the dark charged fermions, its decay width should
be
\begin{eqnarray}
\Gamma_{A'}^{}&=&\Gamma_{A'\rightarrow
\nu\bar{\nu}}^{}+\Gamma_{A'\rightarrow
e\bar{e}}^{}+\Gamma_{A'\rightarrow
u\bar{u}}^{}+\Gamma_{A'\rightarrow
d\bar{d}}^{}\nonumber\\
&=&\frac{5\,\epsilon^2_{}\alpha\cos^2_{}\theta'^{}_W
}{9\,\cos^2_{}\theta^{}_W} m_{A'}^{}\,.
\end{eqnarray}
Here and thereafter
\begin{eqnarray}
\theta^{}_W=\arctan\frac{g^{}_1}{g^{}_2}\,,~~\theta'^{}_W=\arctan\frac{g'^{}_1}{g'^{}_2}\,,
\end{eqnarray}
are the Weinberg angles and
\begin{eqnarray}
\alpha=\frac{g^2_2\sin^2_{}\theta^{}_W}{4\pi}\simeq
\frac{1}{137}\,,~~\alpha'=\frac{g'^2_2\sin^2_{}\theta'^{}_W}{4\pi}\,,
\end{eqnarray}
are the fine structure constants in the ordinary and dark sectors.
We then find
\begin{eqnarray}
\tau_{A'}^{}&\simeq&\left(\frac{4\times
10^{-11}_{}}{\epsilon}\right)^2_{}\left(\frac{\cos\theta^{}_W}{\cos\theta'^{}_W
}\right)^2_{}
\left(\frac{100\,\textrm{MeV}}{m_{A'}^{}}\right)\textrm{sec}\,.\nonumber\\
&&
\end{eqnarray}
So, the dark photon $A'$ with a mass $m_{A'}^{}=100\,\textrm{MeV}$
can have a lifetime shorter than $1$ second if we take $\epsilon>
4\times 10^{-11}_{}$. Currently, the measurement on the muon
magnetic moment constrains $\epsilon^2 \cos^2_{}\theta^{}_W
\cos^2_{}\theta'^{}_W/(1-\epsilon^2_{})< 2\times 10^{-5}_{}$ for
$m_{A'}^{}=100\,\textrm{MeV}$ \cite{pospelov2008}.

From the dark Higgs scalar $\delta'$, which is responsible for the
$U(1)'^{}_{em}$ symmetry breaking, we will have a dark Higgs boson
$h_{\delta'}^{}$ with the mass about
$m_{h_{\delta'}^{}}^{}=2\sqrt{\lambda_{\delta}^{}}\langle\delta'\rangle$.
This dark Higgs boson can mostly decay into the dark photon $A'$
with the decay width,
\begin{eqnarray}
\Gamma_{h_{\delta'}^{}\rightarrow
A'A'}^{}&=&128\,\pi\,\alpha'^2_{}\frac{\langle\delta'\rangle^2_{}}{m_{h_{\delta'}^{}}^{}}
\left[1+\frac{(m_{h_{\delta'}^{}}^2-2m_{A'}^2)^2_{}}{8m_{A'}^4}\right]\nonumber\\
&&\times\sqrt{1-\frac{4m_{A'}^2}{m_{h_{\delta'}^{}}^2}}\,.
\end{eqnarray}
The lifetime $\tau_{h_{\delta'}^{}}^{}$ can be much shorter than $1$
second. For example, we take $\lambda_\delta^{}=1$ and
$\langle\delta'\rangle=117\,\textrm{MeV}$ to give
$m_{h_{\delta'}^{}}^{}=234\,\textrm{MeV}$ and then
\begin{eqnarray}
\tau_{h_{\delta'}^{}}^{}=\frac{1}{\Gamma_{h_{\delta'}^{}}^{}}&=&2.9\times
10^{-22}_{}\left(\frac{\alpha}{\alpha'}\right)^2_{}\,\textrm{sec}\nonumber\\
&& \textrm{for}~~m_{A'}^{}=100\,\textrm{MeV}\,.
\end{eqnarray}

The dark charged leptons are the quasi-Dirac fermions so that their
lepton asymmetries cannot survive \cite{bp2011}. The lightest dark
charged lepton (the dark electron $e'$) thus should have a thermally
produced relic density, which is determined by its pair annihilation
into the dark photon $A'$,
\begin{eqnarray}
\langle\sigma_{e'^{+}_{}e'^{-}_{}\rightarrow
A'A'}^{}v_{\textrm{rel}}^{}\rangle&\simeq& \frac{\pi
\alpha'^2_{}}{m_{f'}^2}\nonumber\\
&=&10^6_{}\,\textrm{pb}\left(\frac{\alpha'}{\alpha}\right)^2_{}
\left(\frac{256\,\textrm{MeV}}{m_{e'}^{}}\right)^2_{}\,.
\end{eqnarray}
It is easy to check the dark electron will have a frozen temperature
far below its mass. This means the dark electron will only give a
negligible contribution to the dark matter relic density since its
number density at the frozen temperature is highly suppressed by a
Boltzmann factor. Similarly, the dark down quark $d'$ and the dark
up quark $u'$ with the masses of the order of GeV will also have the
frozen temperatures far below their masses.

The Big-Bang Nucleosynthesis (BBN) stringently restricts the
existence of the new relativistic degrees of freedom. The constraint
on the new degrees of freedom is conventionally quoted as $\Delta
N_\nu^{}$, the effective number of additional light neutrinos. The
seven-year WMAP observation has measured \cite{komatsu2010}
\begin{eqnarray}
\Delta N_\nu^{}=1.34^{+0.86}_{-0.88} (68\%\,\textrm{CL})\,.
\end{eqnarray}
We now check the dark left-handed neutrinos $v'^{}_L$ which form the
light Dirac neutrinos with the ordinary left-handed neutrinos
$\nu_L^{}$. We can estimate the decoupling temperature of the dark
neutrinos by \cite{kt1990}
\begin{eqnarray}
\left(\frac{g'^{}_2/\cos\theta'^{}_W
}{g^{}_2/\cos\theta^{}_W}\right)^4_{}\left(\frac{\langle\phi\rangle
}{\langle\phi'\rangle}\right)^4_{}G_F^2 T^5_{}=
H(T)\quad\quad\quad\quad\quad\quad&&\nonumber\\
\Rightarrow T\simeq
8.6\,\textrm{GeV}\left(\frac{g_\ast^{}}{100}\right)^{\frac{1}{6}}_{}
\left(\frac{\langle\phi'\rangle}{500\,\langle\phi\rangle}\right)^{\frac{4}{3}}_{}
\left(\frac{g^{}_2/\cos\theta^{}_W}{g'^{}_2/\cos\theta'^{}_W
}\right)^{\frac{4}{3}}_{}\,,&& \nonumber\\
&&
\end{eqnarray}
with $G^{}_F=1.16637\times 10^{-5}_{}\,\textrm{GeV}^{-2}_{}$ being
the Fermi constant \cite{nakamura2010}. At the temperature $T\sim
8\,\textrm{GeV}$, the relativistic degrees of freedom from the
ordinary sector should be about $g_\ast^{}\sim 80$ \cite{kt1990}
while those from the dark sector (the dark photon $A'$, the dark
Higgs scalar $\delta'$, the dark electron $e'$, the dark down quark
$d'$, the dark up quark $u'$ and the dark gluons) could be about
$g_\ast^{}\sim 44.5$. The temperature of the dark neutrinos at the
BBN epoch $T\sim 1\, \textrm{MeV}$ thus should be \cite{kt1990}
\begin{eqnarray}
\left(\frac{T_{\nu'}^{}}{T}\right)^{4}_{}\sim
\left(\frac{10.75}{80+44.5}\right)^{\frac{4}{3}}_{}\simeq 0.038\,,
\end{eqnarray}
and hence give a negligible contribution to the effective number of
additional light neutrinos, i.e.
\begin{eqnarray}
\Delta
N_\nu^{}=\frac{4}{7}\left(\frac{T_{\nu'}^{}}{T}\right)^{4}_{}\lesssim
0.02\,.
\end{eqnarray}

Through the exchange of the dark photon, the dark matter nucleon can
scatter off the ordinary nucleon. For example, the dark matter
proton has the scattering cross section as below,
\begin{eqnarray}
\sigma_{p'N\rightarrow p'N}&\simeq& \frac{\pi\epsilon^2_{} \alpha
\alpha' c'^2_W}{c^2_W}\frac{\mu_r^2}{m_{A'}^4}
\left[\frac{3Z+(A-Z)}{A}\right]^2_{}\,,\nonumber\\
&\simeq &
10^{-41}_{}\,\textrm{cm}^2_{}\left(\frac{\epsilon}{1.5\times
10^{-7}_{}}\right)^2_{}\left(\frac{\alpha'}{\alpha}\right)
\left(\frac{c'^{}_W}{c^{}_W}\right)^2_{}\nonumber\\
&&\times\left(\frac{\mu_r^{}}{0.833\,\textrm{GeV}}\right)^2_{}
\left(\frac{100\,\textrm{MeV}}{m_{A'}^{}}\right)^4_{}\nonumber\\
&&\times \left[\frac{3Z+(A-Z)}{A}\right]^2_{}\,,
\end{eqnarray}
which can be close to the XENON10 limit \cite{angle2011}. Here $Z$
and $A-Z$ are the numbers of proton and neutron within the target
nucleus, while $\mu_r=\frac{m_{p'}^{} m_N^{}}{m_{p'}^{}+m_N^{}}$ is
the reduced mass. Alternatively, the dark neutron can serve as the
dark matter particle if it is lighter than the dark proton. The
detailed studies can be found in \cite{acmy2009}.

\section{Mirror symmetry}

There are many parameters in our models. To reduce the parameters,
we can impose a mirror
\cite{ly1956,kop1966,pavsic1974,bk1982,glashow1986,flv1991,flv1992,abs1992,hodges1993,bdm1996,silagadze1997,cf1998,
mt1999,bcv2001,bb2001,bgg2001,iv2003,fv2003,berezhiani2004,foot2004,berezhiani2005,berezhiani2006,
bb2006-1,bb2006,acmy2009,dlnt2011,chly2012} discrete symmetry under
which the fields transform as
\begin{eqnarray}
&&G'^a_{\mu}\leftrightarrow G^a_{\mu}\,,~W'^a_{\mu}\leftrightarrow
W^a_{\mu}\,,~B'^{}_\mu\leftrightarrow
B^{}_\mu\,,~\phi'\leftrightarrow \phi\,,\nonumber\\
&&\delta'\leftrightarrow\delta\,,~q'^{}_L\leftrightarrow
q^{}_L\,,~d'^{}_R\leftrightarrow d^{}_R\,,~u'^{}_R\leftrightarrow
u^{}_R\,,~l'^{}_L\leftrightarrow
l^{}_L\,,\nonumber\\
&&e'^{}_R\leftrightarrow e^{}_R\,,~N'^{}_R\leftrightarrow
N^{}_R\,,~\Sigma^{}_a\leftrightarrow \Sigma^T_a\,.
\end{eqnarray}
The above mirror symmetry, which is assumed to softly break in the
scalar potential, i.e.
\begin{eqnarray}
\mu^{2}_{\phi'}\neq \mu^{2}_{\phi}\,,~\mu^{2}_{\delta'}\neq
\mu^{2}_{\delta}\,,
\end{eqnarray}
will simplify the parameters to be
\begin{eqnarray}
&&g'^{}_3=g^{}_3\,,~g'^{}_2=g^{}_2\,,~g'^{}_1=g^{}_1\,,~\lambda^{}_{\phi'}=\lambda^{}_{\phi}\,,
~\lambda^{}_{\delta'}=\lambda^{}_{\delta}\,,\nonumber\\
&&\lambda^{}_{\phi\delta}=\lambda^{}_{\phi'\delta'}\,,~\lambda^{}_{\phi\delta'}=\lambda^{}_{\delta\phi'}\,,
~\lambda^{}_{\phi\Sigma}=\lambda^{}_{\phi'\Sigma}\,,
~y_{d'}^{}=y^{}_d\,,\nonumber\\
&&y_{u'}^{}=y^{}_u\,,~y_{e'}^{}=y^{}_e\,,~y_{\delta'}^{}=y^{}_\delta\,,~y_{N'}^{}=y^{}_N\,,
~f^{}_a=f^T_{a}\,,\nonumber\\
&&M^{}_N=M^T_N\,.
\end{eqnarray}
The dark charged fermion masses then should be \cite{bdm1996}
\begin{eqnarray}
\frac{\langle\phi'\rangle}{\langle\phi\rangle}&=&\frac{m_{u'}^{}}{m_{u}^{}}
=\frac{m_{d'}^{}}{m_{d}^{}}=\frac{m_{s'}^{}}{m_{s}^{}}
=\frac{m_{c'}^{}}{m_{c}^{}}=\frac{m_{b'}^{}}{m_{b}^{}}
=\frac{m_{t'}^{}}{m_{t}^{}}\nonumber\\
&=&\frac{m_{e'}^{}}{m_{e}^{}}
=\frac{m_{\mu'}^{}}{m_{\mu}^{}}=\frac{m_{\tau'}^{}}{m_{s\tau}^{}}\,.
\end{eqnarray}

By fixing the VEVs:
\begin{eqnarray}
\langle\phi'\rangle=500\,\langle\phi\rangle\,,
\end{eqnarray}
we can read the dark charged fermion masses \cite{nakamura2010}:
\begin{eqnarray}
\begin{array}{lcrclcr}
m_{d'}^{}&=&2.5\,\textrm{GeV}&\textrm{for}&m_{d}^{}&=&5\,\textrm{MeV}\,,\\
m_{u'}^{}&=&1.25\,\textrm{GeV}&\textrm{for}&m_{u}^{}&=&2.5\,\textrm{MeV}\,,\\
m_{s'}^{}&=&50\,\textrm{GeV}&\textrm{for}&m_s^{}&=&100\,\textrm{MeV}\,,\\
m_{c'}^{}&=&645\,\textrm{GeV}&\textrm{for}&m_{c}^{}&=&1.29\,\textrm{GeV}\,,\\
m_{b'}^{}&=&2.095\,\textrm{TeV}&\textrm{for}&m_{b}^{}&=&4.19\,\textrm{GeV}\,,\\
m_{t'}^{}&=&86.45\,\textrm{TeV}&\textrm{for}&m_{t}^{}&=&172.9\,\textrm{GeV}\,,\\
m_{e'}^{}&=&0.256\,\textrm{GeV}&\textrm{for}&m_{e}^{}&=&0.511\,\textrm{MeV}\,,\\
m_{\mu'}^{}&=&52.85\,\textrm{GeV}&\textrm{for}&m_{\mu}^{}&=&105.7\,\textrm{MeV}\,,\\
m_{\tau'}^{}&=&888.5\,\textrm{GeV}&\textrm{for}&m_{\tau}^{}&=&1.777\,\textrm{GeV}\,.
\end{array}
\end{eqnarray}
The beta functions of the QCD in the ordinary and dark sectors then
will yield
\begin{eqnarray}
\Lambda^{}_{\textrm{QCD}'}&=&\left(\frac{\langle\phi'\rangle}{\langle\phi\rangle}\right)^{\frac{4}{11}}_{}
(m_u^{}m_d^{}m_s^{})^{\frac{2}{33}}_{}\Lambda^{\frac{9}{11}}_{\textrm{QCD}}=1.13\,\textrm{GeV}\nonumber\\
&&\textrm{for}~~\Lambda_{\textrm{QCD}}^{}=200\,\textrm{MeV}\,.
\end{eqnarray}
The dark proton and neutron masses thus can be given by
\begin{eqnarray}
m_{p'}^{}=5\,\textrm{GeV}\,, ~m_{n'}^{}=6.25\,\textrm{GeV}\,.
\end{eqnarray}
In this case, the dark proton will serve as the dark matter
particle.

Under the mirror symmetry, the Dirac neutrino mass matrices
(\ref{idirac}), (\ref{iidirac}) and (\ref{ipiidirac}) should have a
symmetric structure. Compared with the traditional type-I, type-II
or type-I+II Majorana seesaw, our Dirac seesaw will not contain new
parameters since the VEV in the dark sector has been determined by
the dark matter mass.

\section{Summary}

In this paper we have proposed a unified picture of the Dirac
neutrino masses, the baryon asymmetry and the dark matter relic
density. Specifically, we consider an $SU(3)'^{}_c\times
SU(2)'^{}_L\times U(1)'^{}_Y$ dark sector parallel to the
$SU(3)^{}_c\times SU(2)^{}_L\times U(1)^{}_Y$ ordinary sector and
then introduce three types of messenger sectors composed of the
heavy gauge-singlet Dirac fermion(s) and/or $[SU(2)_L^{}\times
SU(2)'^{}_L]$-bidoublet Higgs scalar(s). Through the type-I, type-II
or type-I+II Dirac seesaw mechanism, the heavy fermion singlet(s)
and/or Higgs bidoublet(s) can highly suppress the masses between the
ordinary and dark left-handed neutrinos. So, the ordinary and dark
neutrinos can form the light Dirac neutrinos in a natural way. In
such Dirac seesaw context, the lepton number conserving decays of
the heavy fermion singlet(s) and/or Higgs bidoublet(s) can
simultaneously generate a lepton asymmetry in the
$[SU(2)_L^{}]$-doublet leptons and an opposite lepton asymmetry in
the $[SU(2)'^{}_L]$-doublet leptons. Benefited from the $SU(2)^{}_L$
and $SU(2)'^{}_L$ sphaleron processes, we eventually can obtain a
baryon asymmetry in the ordinary sector and an opposite baryon
asymmetry in the dark sector. The lightest dark nucleon thus should
have a determined mass about $5\,\textrm{GeV}$ to serve as the dark
matter particle. In the presence of the kinetic mixing between the
$U(1)^{}_Y$ and $U(1)'^{}_Y$ gauge fields, the dark nucleons can be
verified in the dark matter direct detection experiments. By
imposing a mirror discrete symmetry, our models needn't more
parameters than the conventional Majorana seesaw models.

\textbf{Acknowledgement}: I thank Shao-Long Chen, Martin Holthausen,
Manfred Lindner and Daniel Schmidt for discussions. This work is
supported by the Sonderforschungsbereich TR 27 of the Deutsche
Forschungsgemeinschaft.

\appendix

\section{The $SU(2)$ and $U(1)$ gauge bosons}

It is easy to find the charged gauge bosons,
\begin{subequations}
\begin{eqnarray}
W^{\pm}_{\mu}&=&\frac{1}{\sqrt{2}}(W^1_{\mu}\mp i
W^2_{\mu})\,,\\
W'^{\pm}_{\mu}&=&\frac{1}{\sqrt{2}}(W'^1_{\mu}\mp i W'^2_{\mu})\,,
\end{eqnarray}
\end{subequations}
and their masses
\begin{subequations}
\begin{eqnarray}
m_{W}^{2}&=&\frac{1}{2}g_2^2(\langle\phi\rangle^2_{}+\sum_a^{}\langle\Sigma_a^{}\rangle^2_{})\,,\\
m_{W'}^{2}&=&\frac{1}{2}g'^{}_2(\langle\phi'\rangle^2_{}+\sum_a^{}\langle\Sigma_a^{}\rangle^2_{})\,.
\end{eqnarray}
\end{subequations}
Furthermore, we can diagonalize the kinetic term of the $U(1)^{}_Y$
and $U(1)'^{}_Y$ gauge fields by making a non-unitary transformation
\cite{fh1991},
\begin{eqnarray}
\tilde{B}^{}_\mu=B^{}_\mu+\epsilon B'^{}_\mu\,,
~~\tilde{B}'^{}_\mu=\sqrt{1-\epsilon^2}B'^{}_\mu\,,
\end{eqnarray}
and then define the orthogonal fields,
\begin{subequations}
\label{orthogonal}
\begin{eqnarray}
&&\left\{\begin{array}{lcl}A^{}_\mu &=&W^3_\mu s^{}_W
+\tilde{B}^{}_\mu c^{}_W\,,\\
[2mm] Z^{}_\mu &=&W^3_\mu c^{}_W
-\tilde{B}^{}_\mu s^{}_W\,;\end{array}\right.\\
[2mm] &&\left\{\begin{array}{lcl}A'^{}_\mu &=&W'^3_\mu s'^{}_{W}
+\tilde{B}'^{}_\mu c'^{}_{W}\,,\\
[2mm] Z'^{}_\mu &=&W'^3_\mu c'^{}_{W} -\tilde{B}'^{}_\mu
s'^{}_{W}\,,\end{array}\right.
\end{eqnarray}
\end{subequations}
with the ordinary and dark Weinberg angles,
\begin{subequations}
\begin{eqnarray}
&&\left\{\begin{array}{l}s^{}_W=\sin\theta^{}_W=\frac{g^{}_1}{\sqrt{g^2_1+g^2_2}}\,,\\
[4mm]
c^{}_W=\cos\theta^{}_W=\frac{g^{}_2}{\sqrt{g^2_1+g^2_2}}\,;\end{array}\right.\\
&&\left\{\begin{array}{l}s'^{}_{W}=\sin\theta'^{}_W=\frac{g'^{}_1}{\sqrt{g'^2_1+g'^2_2}}\,,\\
[2mm]
c'^{}_{W}=\cos\theta'^{}_W=\frac{g'^{}_2}{\sqrt{g'^2_1+g'^2_2}}\,.\end{array}\right.
\end{eqnarray}
\end{subequations}
Among the orthogonal fields (\ref{orthogonal}), $A^{}_\mu$ as the
ordinary photon is exactly massless, while $Z^{}_\mu$, $Z'^{}_\mu$
and $A'^{}_\mu$ have the mass terms as below,
\begin{eqnarray}
\label{zamass} \mathcal{L} &\supset&
\frac{1}{2}m^2_{Z}[Z_\mu^{}+\xi_1^{}s_W^{}(c'^{}_W
A'^{}_\mu-s'^{}_W Z'^{}_\mu)]^2_{}\nonumber\\
&&+\frac{1}{2}m^2_{Z'}[Z'^{}_\mu+\xi_2^{}s'^{}_W(c'^{}_W
A'^{}_\mu-s'^{}_W Z'^{}_\mu)]^2_{}\nonumber\\
&&+\frac{1}{2}m^2_{A'}\left(A'^{}_\mu-\frac{s'^{}_W}{c'^{}_W}Z'^{}_\mu\right)^2_{}\nonumber\\
&&+\frac{1}{2}\xi_3^{}m^2_{Z}\{Z^{}_\mu-[\xi_1^{}s^{}_W
s'^{}_W+c(1-\xi_2^{}s'^2_W)]Z'^{}_\mu\nonumber\\
&&+(\xi_1^{}s^{}_W-c\xi_2^{}s'^{}_W)c'^{}_W A'^{}_\mu\}^2_{}\,.
\end{eqnarray}
Here we have denoted
\begin{eqnarray}
\label{gbmass} m^2_{Z}&=&\frac{g^2_2}{2c^2_W}\langle\phi\rangle^2_{}\,,\nonumber\\
m^2_{Z'}&=&\frac{g'^2_2}{2c'^2_W}\langle\phi'\rangle^2_{}\,,\nonumber\\
m^2_{A'}&=&\frac{8g'^{2}_2
s'^2_{W}}{1-\epsilon^2_{}}\langle\delta'\rangle^2_{}\nonumber\\
&\simeq&\frac{(100\,\textrm{MeV})^2_{}}{1-\epsilon^2_{}}
\left(\frac{\alpha'}{\alpha}\right)\left(\frac{\langle\delta'\rangle}{117\,\textrm{MeV}}\right)^2_{}\,,
\end{eqnarray}
and
\begin{eqnarray}
&&\xi_1^{}=\frac{\epsilon}{\sqrt{1-\epsilon^2}}\simeq
\epsilon\,,~~\xi_2^{}=1-\frac{1}{\sqrt{1-\epsilon^2}}
\simeq -\frac{1}{2}\epsilon^2_{}\,,\nonumber\\
&&\xi_3^{}=\frac{\sum_{a}^{}\langle\Sigma_a^{}\rangle^2_{}}
{\langle\phi\rangle^2_{}}\,,
~~c=\frac{g'^{}_2/c'^{}_W}{g^{}_2/c^{}_W}\,.
\end{eqnarray}
We can further define
\begin{subequations}
\label{physboson}
\begin{eqnarray}
\hat{Z}'^{}_\mu&=&Z'^{}_\mu c_\alpha^{}+ A'^{}_\mu s_\alpha^{}\,,\\
\hat{Z}^{}_\mu&=&Z^{}_\mu c_\beta^{} + (-Z'^{}_\mu s_\alpha^{}+
A'^{}_\mu c_\alpha^{})s_\beta^{}\,,\\
\hat{A}'^{}_\mu&=&-Z^{}_\mu s_\beta^{} + (-Z'^{}_\mu s_\alpha^{}+
A'^{}_\mu c_\alpha^{})c_\beta^{}\,,
\end{eqnarray}
\end{subequations}
with the rotation angles,
\begin{subequations}
\begin{eqnarray}
&&\left\{\begin{array}{l}
c_\alpha^{}=\cos\alpha=\frac{1-\xi_2^{}s'^{2}_W}{\sqrt{(1-\xi_2^{}s'^{2}_W)^2_{}+\xi_2^2s'^2_Wc'^2_W}}\,,\\
[5mm]
s_\alpha^{}=\sin\alpha=\frac{\xi_2^{}s'^{}_Wc'^{}_W}{\sqrt{(1-\xi_2^{}s'^{2}_W)^2_{}+\xi_2^2s'^2_Wc'^2_W}}\,,
\end{array}\right.\\
[5mm] &&\left\{\begin{array}{l}
c_\beta^{}=\cos\beta=\frac{1}{\sqrt{1+\xi_1^2s_W^2(c'^{}_Wc^{}_\alpha+s'^{}_Ws^{}_\alpha)^2_{}}}\,,\\
[5mm]
s_\beta^{}=\sin\beta=\frac{\xi_1^{}s_W^{}(c'^{}_Wc^{}_\alpha+s'^{}_Ws^{}_\alpha)}
{\sqrt{1+\xi_1^2s_W^2(c'^{}_Wc^{}_\alpha+s'^{}_Ws^{}_\alpha)^2_{}}}\,,
\end{array}\right.
\end{eqnarray}
\end{subequations}
to rewrite the mass terms (\ref{zamass}) by
\begin{eqnarray}
\mathcal{L}
&\supset&\frac{1}{2}m_{Z'}^2\left[(1-\xi_2^{}s'^{2}_W)^2_{}+\xi_2^2s'^2_Wc'^2_W\right]\hat{Z}'^2_\mu\nonumber\\
&&+\frac{1}{2}m^2_{Z}\left[\sqrt{1+\xi_1^2s_W^2(c'^{}_Wc_\alpha^{}+s'^{}_Ws_\alpha)^2_{}}\hat{Z}_\mu^{}\right.\nonumber\\
&&\left.+\xi_1^{}s^{}_W(c'^{}_Ws_\alpha-s'^{}_Wc_\alpha^{})\hat{Z}'^{}_\mu\right]^2_{}\nonumber\\
&&+\frac{1}{2}\xi_3^{}m^2_{Z}
\left\{\sqrt{1+\xi_1^2s_W^2(c'^{}_Wc_\alpha^{}+s'^{}_Ws_\alpha)^2_{}}\hat{Z}_\mu^{}\right.\nonumber\\
&&\left.+\left[\xi_1^{}s_W^{}(c'^{}_Ws^{}_\alpha-s'^{}_Wc^{}_\alpha)\right.\right.\nonumber\\
&&\left.\left.-c(1-2\xi_2^{}s'^{}_W+\xi^2_2s'^2_W)\right]\hat{Z}'^{}_\mu\right\}^2_{}\nonumber\\
&&+\frac{1}{2}m_{A'}^2
\left[\left(c_\alpha^{}+\frac{s'^{}_W}{c'^{}_W}s_\alpha^{}\right)(\hat{A}'^{}_\mu c_\beta^{}+\hat{Z}^{}_\mu s_\beta^{})\right.\nonumber\\
&&\left.+\left(s_\alpha^{}-\frac{s'^{}_W}{c'^{}_W}c_\alpha^{}\right)\hat{Z}'^{}_\mu\right]
\end{eqnarray}
Clearly, the definition ({\ref{physboson}}) can give us a physical
dark photon in the case that the dark electromagnetic symmetry is
unbroken. For $\langle\phi'\rangle\gg \langle\phi\rangle\simeq
174\,\textrm{GeV}\gg\langle\delta'\rangle=\mathcal{O}(\textrm{100\,MeV})\gg
\langle\Sigma_a^{}\rangle=\mathcal{O}(\textrm{eV})$ and $\epsilon\ll
1$, the orthogonal fields $Z_\mu^{}$, $Z'^{}_\mu$ and $A'^{}_\mu$
can approximate to the mass eigenstates. The dark photon $A'^{}_\mu$
can couple to both of the dark and ordinary fermions,
\begin{eqnarray} \mathcal{L}&\supset&
\frac{\epsilon ec'^{}_W}{4c_W^{}}A'^{}_\mu\left[\bar{e}\gamma^\mu_{}
\left(3+\gamma_5^{}\right)e+\bar{\nu}\gamma^\mu_{}(1-\gamma_5^{})
\nu\right.\nonumber\\
&&\left.+\bar{d}\gamma^\mu_{}
\left(\frac{1}{3}+\gamma_5^{}\right)d-\bar{u}\gamma^\mu_{}
\left(\frac{5}{3}+\gamma_5^{}\right)u\right]\nonumber\\
&&+e'A'^{}_\mu\left(\frac{1}{3}\bar{d}'\gamma^\mu_{}d'-\frac{2}{3}\bar{u}'\gamma^\mu_{}u'
+\bar{e}'\gamma^\mu_{}e'\right)\nonumber\\
&&\textrm{with} \quad e=g^{}_2 s^{}_W\,,~~e'=g'^{}_2 s'^{}_W\,.
\end{eqnarray}

\section{The dark QCD scale}

The running of the ordinary QCD gauge coupling $\alpha_s^{}(\mu)$ is
given by
\begin{eqnarray}
\alpha_s^{}(\mu)=\frac{2\pi}{11-\frac{2}{3}N_f^{}}\frac{1}{\ln\left[\mu/\Lambda_{\left(N_f^{}\right)}^{}\right]}\,,
\end{eqnarray}
where $N_f^{}$ counts the number of the ordinary quarks involved at
a given scale $\mu$. By matching $\alpha_s^{}(\mu)$ at the scale
$\mu=m^{}_t$ with $N^{}_f=6$ and $N^{}_f=5$, at the scale
$\mu=m^{}_b$ with $N^{}_f=5$ and $N^{}_f=4$, at the scale
$\mu=m^{}_c$ with $N^{}_f=4$ and $N^{}_f=3$, respectively, we can
deduce
\begin{eqnarray}
\label{qcd}
\Lambda_{(5)}^{}&=&m_t^{\frac{2}{23}}\Lambda_{(6)}^{\frac{21}{23}}\,,
~\Lambda_{(4)}^{}=m_b^{\frac{2}{25}}\Lambda_{(5)}^{\frac{23}{25}}\,,\nonumber\\
\Lambda_{(3)}^{}&=&m_c^{\frac{2}{27}}\Lambda_{(4)}^{\frac{25}{27}}\,,
\end{eqnarray}
and then
\begin{eqnarray}
\Lambda_{\textrm{QCD}}^{}=\Lambda_{(3)}^{}=(m_c^{}m_b^{}m_t^{})^{\frac{2}{27}}_{}\Lambda_{(6)}^{\frac{21}{27}}\,.
\end{eqnarray}
Similarly, the dark QCD gauge coupling $\alpha'^{}_s$ should behave
as
\begin{eqnarray}
\alpha'^{}_s(\mu)=\frac{2\pi}{11-\frac{2}{3}N_{f'}^{}}\frac{1}{\ln\left[\mu/\Lambda'^{}_{\left(N_{f'}\right)}\right]}\,,
\end{eqnarray}
with $N_{f'}^{}$ being the number of the involved dark quarks. For
$m_{u'}^{}<m_{d'}^{}<m_{s'}^{}<m_{c'}^{}<m_{b'}^{}<m_{t'}^{}$, we
can have
\begin{eqnarray}
\label{qcd'}
\Lambda'^{}_{(5)}&=&m_{t'}^{\frac{2}{23}}\Lambda'^{\frac{21}{23}}_{(6)}\,,
~\Lambda'^{}_{(4)}=m_{b'}^{\frac{2}{25}}\Lambda'^{\frac{23}{25}}_{(5)}\,,\nonumber\\
\Lambda'^{}_{(3)}&=&m_{c'}^{\frac{2}{27}}\Lambda'^{\frac{25}{27}}_{(4)}\,,
~\Lambda'^{}_{(2)}=m_{s'}^{\frac{2}{29}}\Lambda'^{\frac{27}{29}}_{(4)}\,,\nonumber\\
\Lambda'^{}_{(1)}&=&m_{d'}^{\frac{2}{31}}\Lambda'^{\frac{29}{31}}_{(2)}\,,
~\Lambda'^{}_{(0)}=m_{u'}^{\frac{2}{33}}\Lambda'^{\frac{31}{33}}_{(1)}\,,
\end{eqnarray}
and then
\begin{eqnarray}
\Lambda_{\textrm{QCD}'}^{}=\Lambda'^{}_{(0)}&=&
(m_{u'}^{}m_{d'}^{}m_{s'}^{}m_{c'}^{}m_{b'}^{}m_{t'}^{})^{\frac{2}{33}}_{}
\Lambda'^{\frac{21}{33}}_{(6)}\nonumber\\
&&\textrm{for}~~\Lambda_{\textrm{QCD}'}^{}<m_{u'}^{}\,.
\end{eqnarray}

At the sufficiently high scales $\mu\gg m_{t'}^{}, m_{t}^{}$ the
renormalization group invariants $\Lambda'^{}_{(6)}$ and
$\Lambda^{}_{(6)}$ are only determined by the corresponding strong
gauge couplings $\alpha^{}_s$ and $\alpha'^{}_s$. In the presence of
a mirror symmetry which enforces
\begin{eqnarray}
\alpha^{}_s(\mu)=\alpha'^{}_s(\mu)\Rightarrow
\Lambda'^{}_{(6)}=\Lambda'^{}_{(6)}\,,
\end{eqnarray}
and
\begin{eqnarray}
\frac{\langle\phi'\rangle}{\langle\phi\rangle}=\frac{m_{u'}^{}}{m_{u}^{}}
=\frac{m_{d'}^{}}{m_{d}^{}}=\frac{m_{s'}^{}}{m_{s}^{}}
=\frac{m_{c'}^{}}{m_{c}^{}}=\frac{m_{b'}^{}}{m_{b}^{}}
=\frac{m_{t'}^{}}{m_{t}^{}}\,,
\end{eqnarray}
the dark hadronic scale $\Lambda^{}_{\textrm{QCD}'}$ should arrive
at
\begin{eqnarray}
\Lambda_{\textrm{QCD}'}^{}&=&\left(\frac{\langle\phi'\rangle}{\langle\phi\rangle}\right)^{\frac{4}{11}}_{}
(m_{u}^{}m_{d}^{}m_{s}^{})^{\frac{2}{33}}_{}
\Lambda^{\frac{9}{11}}_{\textrm{QCD}}\nonumber\\
&&\textrm{for}~~\Lambda_{\textrm{QCD}'}^{}<m_{u'}^{}\,.
\end{eqnarray}


\begin{thebibliography}{99}





\bibitem{minkowski1977}
P. Minkowski, Phys. Lett. B \textbf{67}, 421 (1977); T. Yanagida, in
{\it Proceedings of the Workshop on Unified Theory and the Baryon
Number of the Universe}, edited by O. Sawada and A. Sugamoto (KEK,
Tsukuba, 1979), p. 95; M. Gell-Mann, P. Ramond, and R. Slansky, in
{\it Supergravity}, edited by F. van Nieuwenhuizen and D. Freedman
(North Holland, Amsterdam, 1979), p. 315; S.L. Glashow, in {\it
Quarks and Leptons}, edited by M. L\'{e}vy {\it et al.} (Plenum, New
York, 1980), p. 707; R.N. Mohapatra and G. Senjanovi\'{c}, Phys.
Rev. Lett. \textbf{44}, 912 (1980).


\bibitem{mw1980}
M. Magg and C. Wetterich, Phys. Lett. B \textbf{94}, 61 (1980); J.
Schechter and J.W.F. Valle, Phys. Rev. D \textbf{22}, 2227 (1980);
T.P. Cheng and L.F. Li, Phys. Rev. D \textbf{22}, 2860 (1980); G.
Lazarides, Q. Shafi, and C. Wetterich, Nucl. Phys. B \textbf{181},
287 (1981); R.N. Mohapatra and G. Senjanovi\'{c}, Phys. Rev. D
\textbf{23}, 165 (1981).


\bibitem{flhj1989}
R. Foot, H. Lew, X.G. He, and G.C. Joshi, Z. Phys. C \textbf{44},
441 (1989).



\bibitem{rw1983}
M. Roncadelli and D. Wyler, Phys. Lett. B \textbf{133}, 325 (1983);
P. Roy and O. Shanker, Phys. Rev. Lett. \textbf{52}, 713 (1984).


\bibitem{mp2002}
H. Murayama and A. Pierce, Phys. Rev. Lett. \textbf{89}, 271601
(2002).

\bibitem{gh2006}
P.H. Gu and H.J. He, JCAP \textbf{0612}, 010 (2006).



\bibitem{fy1986}
M. Fukugita and T. Yanagida, Phys. Lett. B \textbf{174}, 45 (1986).

\bibitem{mz1992}
R.N. Mohapatra and X. Zhang, Phys. Rev. D \textbf{46}, 5331 (1992);
E. Ma and U. Sarkar, Phys. Rev. Lett. \textbf{80}, 5716 (1998).

\bibitem{fps1995}
M. Flanz, E.A. Paschos, and U. Sarkar, Phys. Lett. B \textbf{345},
248 (1995); M. Flanz, E.A. Paschos, U. Sarkar, and J. Weiss, Phys.
Lett. B \textbf{389}, 693 (1996); L. Covi, E. Roulet, and F.
Vissani, Phys. Lett. B \textbf{384}, 169 (1996); A. Pilaftsis, Phys.
Rev. D \textbf{56}, 5431 (1997).

\bibitem{hms2000}
T. Hambye, E. Ma, and U. Sarkar, Nucl. Phys. B \textbf{602}, 23
(2001).


\bibitem{di2002}
S. Davidson and A. Ibarra, Phys. Lett. B \textbf{535}, 25 (2002); W.
Buchm\"{u}ller, P. Di Bari, and M. Pl\"{u}macher, Nucl. Phys. B
\textbf{665}, 445 (2003).


\bibitem{hs2004}
T. Hambye and G. Senjanovi\'{c}, Phys. Lett. B \textbf{582}, 73
(2004); S. Antusch and S.F. King, Phys. Lett. B \textbf{597}, 199
(2004).

\bibitem{hlnps2004}
T. Hambye, Y. Lin, A. Notari, M. Papucci, A. Strumia, Nucl. Phys. B
\textbf{695}, 169 (2004).


\bibitem{hrs2005}
T. Hambye, M. Raidal, and A. Strumia, Phys. Lett. B \textbf{632},
667 (2006).


\bibitem{dnn2008}
S. Davidson, E. Nardi, and Y. Nir, Phys. Rept. \textbf{466}, 105
(2008).


\bibitem{bd2009}
S. Blanchet and P. Di Bari, Nucl. Phys. B \textbf{807}, 155 (2009).




\bibitem{gu2012}
P.H. Gu, Phys. Lett. B \textbf{713}, 485 (2012).

\bibitem{dlrw1999}
K. Dick, M. Lindner, M. Ratz, and D. Wright, Phys. Rev. Lett.
\textbf{84}, 4039 (2000).

\bibitem{tt2006}
B. Thomas and M. Toharia, Phys. Rev. D \textbf{73}, 063512 (2006);
S. Abel and V. Page, JHEP \textbf{0605}, 024 (2006); D.G. Cerdeno,
A. Dedes, and T.E.J. Underwood, JHEP \textbf{0609}, 067 (2006); E.J.
Chun and P. Roy, JHEP \textbf{0806}, 089 (2008); P.H. Gu, H.J. He,
and U. Sarkar, Phys. Lett. B \textbf{659}, 634 (2008); A. Bechinger
and G. Seidl, Phys. Rev. D \textbf{81}, 065015 (2010); H. Davoudiasl
and I. Lewis, arXiv:1112.1939 [hep-ph].




\bibitem{komatsu2010}
E. Komatsu {\it et al.}, Astrophys. J. Suppl. \textbf{192}, 18
(2011).








\bibitem{ly1956}
T.D. Lee and C.N. Yang, Phys. Rev. textbf{104}, 254 (1956).


\bibitem{kop1966}
I.Yu. Kobzarev, L.B. Okun, and I.Ya. Pomeranchuk, Sov. J. Nucl.
Phys. \textbf{3}, 837 (1966) [Yad. Fiz. \textbf{3}, 1154 (1966)].

\bibitem{pavsic1974}
M. Pavsic, Int. J. Theor. Phys. \textbf{9}, 229 (1974).


\bibitem{bk1982}
S.I. Blinnikov and M.Y. Khlopov, Sov. J. Nucl. Phys. \textbf{36},
472 (1982) [Yad. Fiz. \textbf{36}, 809 (1982)]; S.I. Blinnikov and
M.Y. Khlopov, Sov. Astron. \textbf{27}, 371 (1983) [Astro. Zh.
\textbf{60}, 632 (1983)].


\bibitem{glashow1986}
S. L. Glashow, Phys. Lett. B \textbf{167}, 35 (1986).


\bibitem{flv1991}
R. Foot, H. Lew, and R.R. Volkas, Phys. Lett. B \textbf{272}, 67
(1991).

\bibitem{flv1992}
R. Foot, H. Lew, and R.R. Volkas, Mod. Phys. Lett. A \textbf{7},
2567 (1992); R. Foot and R.R. Volkas, Phys. Rev. D \textbf{52}, 6595
(1995).




\bibitem{abs1992}
E.H. Akhmedov, Z. Berezhiani, and G. Senjanovi\'{c}, Phys. Rev.
Lett. \textbf{69}, 3013 (1992); Z. Berezhiani and R.N. Mohapatra,
Phys. Rev. D \textbf{52}, 6607 (1995).






\bibitem{silagadze1997}
Z. Silagadze, Phys. Atom. Nucl. \textbf{60}, 272 (1997) [Yad. Fiz.
\textbf{60N2}, 336 (1997)].





\bibitem{hodges1993}
H.M. Hodges, Phys. Rev. D \textbf{47} 456 (1993).



\bibitem{bdm1996}
Z. Berezhiani, Acta. Phys. Polon. B \textbf{27}, 1503 (1996); Z.
Berezhiani, A. Dolgov, and R.N. Mohapatra, Phys. Lett. B
\textbf{375}, 26 (1996).



\bibitem{cf1998}
M. Collie and R. Foot, Phys. Lett. B \textbf{432}, 134 (1998); R.
Foot and R.R. Volkas, Phys. Rev. D \textbf{61}, 043507 (2000).



\bibitem{mt1999}
R.N. Mohapatra and V.L. Teplitz, Phys. Lett. B \textbf{462}, 302
(1999); R.N. Mohapatra and V.L. Teplitz, Phys. Rev. D \textbf{62},
063506 (2000); R.N. Mohapatra, S. Nussinov, and V.L. Teplitz, Phys.
Rev. D \textbf{66}, 063002 (2002).






\bibitem{bcv2001}
Z. Berezhiani, D. Comelli, and F.L. Villante, Phys. Lett. B
\textbf{503} 362 (2001); Z. Berezhiani, P. Ciarcelluti, D. Comelli,
and F.L. Villante, Int. J. Mod. Phys. D 14, 107 (2005); P.
Ciarcelluti, Int. J. Mod. Phys. D 14, 187 (2005).



\bibitem{bb2001}
L. Bento and Z. Berezhiani, Phys. Rev. Lett. \textbf{87}, 231404
(2001).


\bibitem{bgg2001}
Z. Berezhiani, L. Gianfagna, and M. Giannotti, Phys. Lett. B
\textbf{500}, 286 (2001); Z. Berezhiani and A. Lepidi, Phys. Lett. B
\textbf{681}, 276 (2009).



\bibitem{iv2003}
A.Y. Ignatiev and R.R. Volkas, Phys. Rev. D \textbf{68}, 023518
(2003).


\bibitem{fv2003}
R. Foot and R.R. Volkas, Phys. Rev. D \textbf{68}, 021304 (2003); R.
Foot and R.R. Volkas, Phys. Rev. D \textbf{69}, 123510 (2004).


\bibitem{berezhiani2004}
L. Bento and Z. Berezhiani, Int. J. Mod. Phys. A \textbf{19}, 3775
(2004)





\bibitem{foot2004}
R. Foot, Phys. Rev. D \textbf{69}, 036001 (2004); R. Foot, Phys.
Rev. D \textbf{74}, 023514 (2006); R. Foot, Phys. Rev. D
\textbf{78}, 043529 (2008); P. Ciarcelluti and R. Foot, Phys. Lett.
B \textbf{679}, 278 (2009); R. Foot, Phys. Rev. D \textbf{81},
087302 (2010); R. Foot, Phys. Rev. D \textbf{86}, 023524 (2012);
Phys. Lett. B \textbf{718}, 745 (2013).




\bibitem{berezhiani2005}
Z. Berezhiani, hep-ph/0508233.

\bibitem{berezhiani2006}
Z. Berezhiani, AIP Conf. Proc. \textbf{878}, 195 (2006); Z.
Berezhiani, Eur. Phys. J. ST \textbf{163}, 271 (2008).




\bibitem{bb2006-1}
L. Bento and Z. Berezhiani, Phys. Rev. Lett. \textbf{96}, 081801
(2006).

\bibitem{bb2006}
L. Bento and Z. Berezhiani, Phys. Lett. B \textbf{635}, 253 (2006).

\bibitem{acmy2009}
H. An, S.L. Chen, R.N. Mohapatra, and Y. Zhang, JHEP \textbf{1003},
124 (2010); H. An, S.L. Chen, R.N. Mohapatra, S. Nussinov, and Y.
Zhang, Phys. Rev. D \textbf{82}, 023533 (2010).


\bibitem{dlnt2011}
C.R. Das, L.V. Laperashvili, H.B. Nielsen, and A. Tureanu, Phys.
Rev. D \textbf{84}, 063510 (2011).

\bibitem{chly2012}
J.W. Cui, H.J. He, L.C. L\"{u}, and F.R. Yin, Phys. Rev. D
\textbf{85}, 096003 (2012).








\bibitem{nussinov1985}
S. Nussinov, Phys. Lett. B \textbf{165}, 55 (1985).

\bibitem{bcf1990}
S.M. Barr, R.S. Chivulula, and E. Farhi, Phys. Lett. B \textbf{241},
387 (1990); S.M. Barr, Phys. Rev. D \textbf{44}, 3062 (1991).

\bibitem{kaplan1992}
D.B. Kaplan, Phys. Rev. Lett. \textbf{68}, 741 (1992).


\bibitem{dgw1992}
S. Dodelson, B.R. Greene, and L.M. Widrow, Nucl. Phys. B
\textbf{372}, 467 (1992).

\bibitem{kuzmin1998}
V.A. Kuzmin, Phys. Part. Nucl. \textbf{29}, 257 (1998) [Fiz. Elem.
Chast. Atom. Yadra \textbf{29}, 637 (1998)].


\bibitem{kl2005}
R. Kitano and I. Low, Phys. Rev. D \textbf{71}, 023510 (2005).


\bibitem{as2005}
K. Agashe and G. Servant, JCAP \textbf{0502}, 002 (2005); M.
Cirelli, P. Panci, G. Servant, and G. Zaharijas, JCAP \textbf{1203},
015 (2012).



\bibitem{clt2005}
N. Cosme, L. Lopez Honorez, and M.H.G. Tytgat, Phys. Rev. D
\textbf{72}, 043505 (2005).







\bibitem{gsz2009}
P.H. Gu, U. Sarkar, and X. Zhang, Phys. Rev. D \textbf{80}, 076003
(2009); P.H. Gu and U. Sarkar, Phys. Rev. D \textbf{81}, 033001
(2010); P.H. Gu, M. Lindner, U. Sarkar, and X. Zhang, Phys. Rev. D
\textbf{83}, 055008 (2011).



\bibitem{dmst2010}
H. Davoudiasl, D.E. Morrissey, K. Sigurdson, and S. Tulin, Phys.
Rev. Lett. \textbf{105}, 211304 (2010); H. Davoudiasl, D.E.
Morrissey, K. Sigurdson, and S. Tulin, Phys. Rev. D \textbf{84},
096008 (2011); N. Blinov, D.E. Morrissey, and S. Tulin,
arXiv:1206.3304 [hep-ph].


\bibitem{bdfr2011}
M. Blennow, B. Dasgupta, E. Fernandez-Martinez, and N. Rius, JHEP
\textbf{1103}, 014 (2011); M. Blennow, E. Fernandez-Martinez, J.
Rendondo, and P. Serra, arXiv:1203.5805 [hep-ph].

\bibitem{mcdonald2011}
J. McDonald, Phys. Rev. D \textbf{83}, 083509 (2011); Phys. Rev. D
\textbf{84}, 103514 (2011).






\bibitem{hmw2010}
L.J. Hall, J. March-Russell, and S.M. West, arXiv:1010.0245
[hep-ph]; J. March-Russell, M. McCullough, JCAP \textbf{1203}, 019
(2012); J. March-Russell, J. Unwin, and S.M. West, arXiv:1203.4854
[hep-ph].


\bibitem{dk2011}
B. Dutta and J. Kumar, Phys. Lett. B \textbf{699}, 364 (2011).


\bibitem{frv2011}
A. Falkowski, J.T. Ruderman, and T. Volansky, JHEP \textbf{1105},
106 (2011).






\bibitem{hms2011}
N. Haba, S. Matsumoto, and R. Sato, Phys. Rev. D \textbf{84}, 055016
(2011).


\bibitem{kllly2011}
Z. Kang, J. Li, T. Li, T. Liu, and J. Yang, arXiv:1102.5644
[hep-ph]; Z. Kang and T. Li, arXiv:1111.7313 [hep-ph].


\bibitem{gsv2011}
M.L. Graesser, I.M. Shoemaker, and L. Vecchi, JHEP \textbf{1110},
110 (2011).






\bibitem{fss2011}
M.T. Frandsen, S. Sarkar, and K. Schmidt-Hoberg, Phys. Rev. D
\textbf{84}, 051703 (2011).

\bibitem{myz2012}
S.D. McDermott, H.B. Yu, and K.M. Zurek, Phys. Rev. D \textbf{85},
023519 (2012); S. Tulin, H.B. Yu, and K.M. Zurek, JCAP
\textbf{1205}, 013 (2012).

\bibitem{idc2011}
H. Iminniyaz, M. Drees, and X. Chen, JCAP \textbf{107}, 003 (2011).



\bibitem{bpsv2011}
N.F. Bell, K. Petraki, I.M. Shoemaker, and R.R. Volkas, Phys. Rev. D
\textbf{84}, 123505 (2011); K. Petraki, M. Trodden, and R.R. Volkas,
JCAP \textbf{1202}, 044 (2012); B. von Harling, K. Petraki, and R.R.
Volkas, JCAP \textbf{1205}, 021 (2012).



\bibitem{cr2011}
Y. Cui, L. Randall, and B. Shuve, JHEP \textbf{1108}, 073 (2011); Y.
Cui, L. Randall, and B. Shuve, JHEP \textbf{1204}, 075 (2012).

\bibitem{as2012}
C. Arina, and N. Sahu, Nucl. Phys. B \textbf{854}, 666 (2012); C.
Arina, J.O. Gong, and N. Sahu, arXiv: 1206.0009 [hep-ph].




\bibitem{ms2011}
E. Ma and U. Sarkar, arXiv:1111.5350 [hep-ph].



\bibitem{dm2012}
H. Davoudiasl and R.N. Mohapatra, arXiv:1203.1247 [hep-ph].

\bibitem{fnp2012}
W.Z. Feng, P. Nath, and G. Peim, arXiv:1204.5752 [hep-ph].








\bibitem{thooft1976}
G. t'Hooft, Phys. Rev. Lett. \textbf{37}, 8 (1976); Phys. Rev. D
\textbf{14}, 3432 (1976).


\bibitem{krs1985}
V.A. Kuzmin, V.A. Rubakov, and M.E. Shaposhnikov, Phys. Lett. B
\textbf{155}, 36 (1985).


\bibitem{abs2005}
T. Asaka, S. Blanchet, and M. Shaposhnikov, Phys. Lett. B
\textbf{631}, 151 (2005); A. Boyarsky, O. Ruchayskiy, and M.
Shaposhnikov, Ann. Rev. Nucl. Part. Sci. \textbf{59}, 191 (2009).


\bibitem{ftv2012}
D.V. Forero, M. T\'{o}rtola, and J.W.F. Valle, Phys. Rev. D
\textbf{86}, 073012 (2012).


\bibitem{bp2011}
M.R. Buckley and S. Profumo, Phys. Rev. Lett. \textbf{108}, 011301
(2012).


\bibitem{kt1990}
E.W. Kolb and M.S. Turner, \textit{The Early Universe},
Addison-Wesley, 1990.






\bibitem{nakamura2010}
K. Nakamura {\it et al.}, (Particle Data Group), J. Phys. G
\textbf{37}, 075021 (2010).





\bibitem{pospelov2008}
M. Pospelov, Phys. Rev. D \textbf{80}, 095002 (2009).

\bibitem{angle2011}
J. Angle {\it et al.}, (XENON10 Collaboration), Phys. Rev. Lett.
\textbf{107}, 051301 (2011).


\bibitem{fh1991}
R. Foot and X.G. He, Phys. Lett. B \textbf{267}, 509 (1991).




\end{thebibliography}
\end{document}